\documentclass[
reprint,
superscriptaddress,
nofootinbib,
 amsmath,amssymb,
 aps, prc,
]{revtex4-2}

\usepackage{graphicx}
\usepackage{dcolumn}
\usepackage{bm}
\usepackage{hyperref}

\begin{document}

\preprint{APS/123-QED}

\title{Extension of the integrated hydrokinetic model to nuclear collision energies relevant for the RHIC Beam-Energy Scan program and the research program at GSI-FAIR}


\author{Musfer Adzhymambetov}
\email{adzhymambetov@gmail.com}
\affiliation{
Bogolyubov Institute for Theoretical Physics,
Metrolohichna  14b, 03143 Kyiv,  Ukraine 
}

\author{Yuri Sinyukov}%
\email{sinyukov@bitp.kyiv.ua}
\affiliation{
Bogolyubov Institute for Theoretical Physics,
Metrolohichna  14b, 03143 Kyiv,  Ukraine 
}
\affiliation{
Faculty of Physics, Warsaw University of Technology,
75 Koszykowa street, 00-662, Warsaw, Poland 
}
\date{\today}
\begin{abstract}

The present work is devoted to developing the integrated hydrokinetic model (iHKM) for relativistic nucleus-nucleus collisions. While the previous cycle of works on this topic focused on ultra-relativistic collisions at the top RHIC and different LHC energies, the current work addresses relativistic collisions at the lower energies, specifically ranging from approximately 2 to 50 GeV per nucleon pair in the center-of-mass colliding system.  In such collisions, the formation times for the initial state of dense matter can be up to three orders of magnitude longer than those in ultra-relativistic collisions. This difference reflects a fundamentally distinct nature and formation process, particularly regarding the possible stages of initial state evolution, including thermalization (which may be only partial at very low collision energies), subsequent hydrodynamic expansion, and the final transition of matter evolution into a hadronic cascade.  These stages, which are fully realized in ultra-relativistic reactions, can also occur within the energy range of BES RHIC, albeit with distinct time scales. This publication not only advances the theoretical development of iHKM (referred to, if necessary, as the {\it{extended}} version of integrated Hydrokinetic Model, iHKM{\it{e}}), but also provides examples of model applications for calculating observables. A systematic description across a wide range of experimental energies, which is preliminary yet quite satisfactory, for spectra, flow, and femtoscopy, will follow this study.
\end{abstract}

\maketitle

\section{\label{sec:level1}  Introduction} 
The goal of relativistic heavy ion experiments, which have been continuing for more than thirty years at accelerators/colliders of different generations - from the AGS to the LHC, is to create and study the new forms of strongly interacting matter with extremely high densities and temperatures. The energy density reached in such systems resembles that governing the Early Universe just microseconds after the initial singularity \cite{Rafelski:2013qeu}. Such matter can form at some internal stage of the nuclei collision process when the system created becomes almost thermal, while the energy density in an expanding fireball is still very high \cite{STAR:2005gfr, PHENIX:2004vcz, PHOBOS:2004zne, BRAHMS:2004adc}. 

In the 1950s, the idea to describe the proton-proton and nuclear-nuclear collision processes of multi-particle production in the models of hydrodynamics type appeared. This new tendency, as to compare with S-matrix formalism \cite{Wheeler:1937zz, Collins:1977jy, Barone:2002cv}, has been started from a pure hydrodynamic model, called now the Landau model, where the simplest prescription for initial and final matter states in the collisions of particles/nuclei \cite{Landau:1953gs} have been used. Later, in the 80th,  further development of the models goes through the so-called Bjorken model \cite{Bjorken:1982qr} and hydrodynamically inspired parametrizations for the final hydro-collision stage (see,e.g. \cite{Borysova:2005ng}). The consequent development includes all the stages (currently five, more details are below) of the evolution of superdense matter created in heavy-ion collisions. As to the high energy situation,  at the top RHIC and all the LHC energies, the new form of matter - quark-gluon plasma and hadron-resonance gas - manifested itself in the soft physics observables, which include hadron and photon yields,  spectra, and particle correlations. All these measurements are well described in the integrated HydroKinetic Model - iHKM \cite{Naboka:2014eha, Shapoval:2020nec}, which we will try to generalize for intermediate and small relativistic energies in this paper.   
    
Experiments in the intermediate and low relativistic collision energy range are of special interest. The currently acting ones are associated with the Beam Energy Scan program at RHIC (BES RHIC) and the HADES experiment at the GSI accelerator facility. The nearest planning is the Compressed Baryon Matter (CBM) experiment at the GSI-FAIR. Despite collision processes at the LHC, where the transition between hadron and quark-gluon matters is happening without the phase transition (crossover), at the above-mentioned experiments characterized by large net baryon densities in creating matter, one hopes to search for the thermodynamic line of the phase transition between hadron and quark-gluon matters and also try to discover the critical endpoint \cite{Guenther:2020jwe}. 
     
Therefore, this series of experiments on relativistic nucleus-nucleus collisions is the guiding light for the development of effective theoretical models of strongly interacting matter. It is essential to take into account that the new forms of matter arise in the collision processes during only one of the concise stages (which lasts $10^{-23}$ – $10^{-22}$ sec) of ultrafast evolution of the matter, and so a detailed analysis of the properties of its new forms needs the construction of a complete (all-stages) dynamic picture of the collisions.
      
The objective of this paper is to present an extended integrated hydrokinetic model for the soft physics in all mentioned experiments, covering (all together with already developed iHKM for ultra-relativistic energies)  the range from 2 GeV up to 10 TeV energies per nucleon colliding pair in their center of mass, within a unified approach based on the extended integrated HydroKinetic Model - iHKMe.  The latter will supplement the iHKM, which is already available for ultra-relativistic energies, to the intermediate and low relativistic collision energies.  For each considered energy, all the possible stages of nuclear collision processes will be investigated within the same unified description as in the original iHKM. The mentioned stages of the matter evolution during the collision process are: the formation of the initial conditions for the system expansion into a vacuum just after the collision, gradual thermalization of created superdense matter (maybe not complete for quite low collision energies, that can happen without hydro-stage),  its consequent hydrodynamic evolution for intermediate energies, the translation of the description of hydrodynamic/or partially thermal medium to particle language – so-called particlization,  and,  finally, the cascade stage for still interacting already individual particles \cite{Naboka:2015qra}. So, our approach also includes the possibility that at fairly low energies, not all of these stages (e.g., hydro-stage) will be realized/activated. 
   
The description of the soft physics observables within this unified approach allows one to conclude: at which energies the quark-gluon plasma is created, when the phase transition takes place, and whether the critical endpoint occurs at some collision energy. In addition, from the correlation femtoscopy analysis for baryons, it is planned to extract the most important characteristics of their strong interactions, such as scattering lengths and others. 
    
One of the most serious difficulties in describing a complete set of soft physics observables, just combining differently developed stages into a single picture, was that one needs to start hydrodynamics as soon as possible, just after colliding nuclei overlap. Otherwise, if one starts hydrodynamic evolution later, say, at a typical time-scale for strong interactions around  1~fm/$c$,  either spectra, particle correlation functions, or anisotropic flow will describe the experimental data unsatisfactorily. The reason is that at the standard mentioned starting time for viscous hydro-evolution, neither radial nor anisotropic collective flow at the freeze-out stage develops well to describe data,  because of a lack of time for pressure to accelerate enough the created system transversally. Such a logic gives rise to intensive theoretical attempts to explain very early thermalization/hydrodynamization: ADS/CFT correspondence \cite{Aharony:1999ti, Maldacena:2003nj, Maldacena:2011ut, Casalderrey-Solana:2011dxg, Bernamonti:2011vm, DeWolfe:2013cua}, Unruh effect \cite{Unruh:1976db, Crispino:2007eb}, three-gluon production, etc. They continued for almost two decades, but have not been successful.  
    
Our idea, and later its full realization, was: to get more intensive flow, both radial and anisotropic, one does not need the pressure gradient in the created fireball and, so, does not need the early fast thermalization \cite{Sinyukov:2009xe}. These flows can be well developed because only the geometrical form of the very initial system, which is essentially finite in nucleus-nucleus collisions and, in addition, has an anisotropic shape in non-central collisions. What is very important - in any case, one needs the system's thermalization sooner or later to describe the data, but not necessarily a fast one. If it happens, say, at 1 fm/$c$, the transverse flow will be present already at this (relatively late) thermalization/ hydrodynamization time because they were developed at the pre-thermal stage, even with quite gradually appearing pressure. The corresponding formalism developed allows one to build the full model of heavy ion collisions that incorporates all stages of heavy ion collision processes without the physically controversial hypothesis of the very early system’s thermalization \cite{Sinyukov:2009xe, Naboka:2015qra}. 

This paper is developing the iHKM model for the intermediate energy range of relativistic nucleus-nucleus collisions: programs BES at RHIC,  and future  CBM in  GSI-FAIR experiments, including current HADES activity.  We will call the corresponding new extension of the model in this paper - iHKMe, or again iHKM, when it is clear for which energy it is applied. The main modification in the newly developed model concerns the formation of initial conditions of matter evolution in relativistic A+A collisions at relatively small energies. In the already built iHKM, the overlapping time of colliding nuclei is around  $10^{-3}$ fm/c at the energy 5.02~ATeV. At BES RHIC energy, e.g., 14.5 AGeV, it is 1.6~fm/$c$.  
           
It is clear that the model of initial conditions for the consequent pre-thermal (thermalization) stage is very different. While in iHKM we use a hybrid approach based on MC-GLAUBER calculations, realized in the GLISSANDO-2 model \cite{Rybczynski:2013yba} in the transverse direction, and Color Glass Condensate in the longitudinal one, the initial conditions are dramatically changed in the collisions at BES RHIC and below energies where the overlapping time is differed (larger) by 3 order of the value. The initial conditions in the extended for intermediate and small relativistic heavy ion collision energies model - iHKMe are based on the quasiclassical UrQMD simulations \cite{Bleicher:1999xi} with added quantum (and classical) fluctuations during the thermalization process. It leads to partial thermal evolution of the matter, created at small and intermediate collision energies.

The goal of this paper is to propose the theoretical basis for the description of the soft physics at the intermediate and small relativistic energies, in addition to what has been done already for ultrarelativistic energies (iHKM), and also to illustrate the results of the developed model at one of the collision energies. A consistent description of the observables within iHKMe/iHKM in BES RHIC, and GSI-FAIR experiments will be presented in subsequent publications.

\section{Model description}

\subsection{The basic aspects of the approach}
\label{subsec:IIA}

In the extended version of iHKM, hereafter referred to as iHKMe, when necessary, the initial pre-equilibrium dynamics at low and intermediate relativistic collision energies are modeled using the UrQMD hadron-string cascade approach~\cite{Bleicher:1999xi}. This approach was first employed as an initial stage for hydrodynamic simulations in Ref.~\cite{Petersen:2008dd}. It offers several advantages: it describes non-equilibrium dynamics and provides event-by-event fluctuating, $3+1$ dimensional distributions of energy, momentum, and conserved charge densities (baryon, electric, and strangeness), while explicitly conserving these quantities. Notably, it enables the use of the same microscopic model for both the initial and final (afterburner) stages, allowing for a continuous description of the system’s peripheral regions (the corona), where thermalization is not achieved. Our model architecture also allows the incorporation of other similar transport approaches, such as SMASH~\cite{Petersen:2018jag}, JAM~\cite{Nara:1999dz}, PHSD~\cite{Cassing:2009vt}, and GiBUU~\cite{Buss:2011mx}. This includes variants with density-dependent potentials that influence the equation of state~\cite{Nara:2021fuu, SMASH:2016zqf}, which may be particularly relevant at a few GeV collision energies due to the importance of the initial compression stage, as demonstrated in Ref.~\cite{OmanaKuttan:2022the}. However, these alternatives are not currently implemented.

The transformation between the hydrodynamic description and the microscopic particle evolution presents significant challenges, particularly at relatively low but still relativistic collision energies, such as $\sqrt{s_{NN}} < 10$~GeV in $A+A$ collisions. The underlying reasons for this complexity are the following. First, a fundamental difficulty arises because during UrQMD (or other transport model) simulations of the initial collision stage, the gradual formation of a hydrodynamic subsystem from the total system cannot be based on the initial singular particle distribution from a single cascade event. Therefore, some smoothing or averaging procedure is required even in event-by-event analyses (a comprehensive review of this topic is provided in Ref.~\cite{Oliinychenko:2015lva}).

Second, the system may not become fully hydrodynamized at very low collision energies. In such cases, the hydrodynamically expanding part of the matter transforms into final-state particles within a surrounding region of the interacting hadron gas that never entered the hydrodynamic phase during the system's evolution. The process by which the hydrodynamic matter converts into hadrons is referred to as \emph{particlization}, and it serves as the starting point for the subsequent hadronic cascade stage in the total system. A central theoretical challenge in this framework is to describe the physical mechanism underlying thermalization or hydrodynamization. The approach employed in Ref.~\cite{Naboka:2015qra}, which is currently used in our model, provides a phenomenological description based on the Boltzmann equation with relaxation and thermalization times. The gradual transformation of initially non-thermal matter into (partially) thermalized matter is governed by conservation law equations. Other attempts to model the gradual fluidization of nonequilibrium matter have been proposed. For example, in Ref.~\cite{Shen:2017bsr}, strings are dynamically formed between participant quarks or nucleons and subsequently thermalize, serving as a source term for the hydrodynamic evolution. A similar mechanism of continuous fluidization from strings or hadronic resonances has been introduced in the context of the UrQMD and JAM transport models in Refs.~\cite{Akamatsu:2018olk, Du:2018mpf}, respectively.

Third, the important question concerns the nature of the thermal fluid that appears against the backdrop of the initial gas of colliding hadrons. We suppose that initially, it appears due to local quantum and classical density fluctuations through the QCD droplet formation. Similar to our first point (see above), even at event-by-event analysis, the smoothing procedure is necessary to describe some stage of the collision process in the continuous medium approximation. We just smooth out the droplet picture for very high-density local fluctuations into an effectively hydrodynamic one. 

Fourth, the question arises about the particlization of the fluid component. At not very low relativistic energies, at some stage of the matter evolution, the droplets (even without any smearing) can merge into a liquid, and this substance fills almost the entire system. Then the description of the particlization process has a standard form {\it a la} Cooper-Frey prescription. However, at low energies, density fluctuations that form droplets may be rare and, accordingly, the corresponding hydrodynamically averaged component is small (or even absent), and a significant part of the system consists of the UrQMD-particle component at the hypersurface of particlization. This possible scenario should also be developed in the iHKMe approach, designed to describe nuclear collisions at energies from 1-3 to 40-50 AGeV. 

Let us consider the particlization and its condition in some detail. We start with a pure hydrodynamic system. A near-local thermal equilibrium and hydrodynamic behavior can be maintained in a finite expanding system as long as the collision rate among the particles is much faster than the expansion rate. Since densities drop out during 3-dim expansion intensively, the collision rate decreases rapidly, and the system eventually falls out of equilibrium. As a result, the hydrodynamic medium decouples, and freeze-out or the hadron cascade phase happens. 

The inverse expansion rate is the collective expansion time scale \cite{Akkelin:2001wv}:
\begin{equation}
\tau _{\exp }=-\left( \frac{1}{n}\frac{\partial n}{\partial t^{\ast }}%
\right) ^{-1}=-\left( \frac{1}{n}u^{\mu}\partial _{\mu }n\right) ^{-1}=%
\frac{1}{\partial_{\mu }u^{\mu }},  \label{expansion-time}
\end{equation}
where $t^{\ast }$ is the proper time in the fluid's local rest system. The last approximate equality follows from the conservation of particle number density currents at the last stage of the hydro-expansion. 

The inverse scattering rate of particle species $i$ is the mean time between scattering events for particle $i$, $\tau _{scat}^{(i)}$:
\begin{equation}
\tau _{scat}^{(i)}\approx \frac{1}{\sum \left\langle v_{ij}\sigma
_{ij}\right\rangle n_{j}},  \label{scattering-time}
\end{equation}
where $v_{ij}$ is the relative velocity between the scattering particles and $\sigma _{ij}$ is the total cross section between particles $i$ and $j$, the sharp brackets mean an average over the local thermal distributions. This time is determined by the densities of all particles with which particle $i$ can scatter, and the corresponding scattering cross sections. Let us estimate the mean time between scatterings for pions, $\tau_{scat}^{(\pi )}$. First, note that $\tau_{scat}^{(i)}>\lambda ^{(i)}/c$, where $\lambda^{(i)} $ is mean free path for particle species $i$, $c=1$ is  the light velocity:
\begin{equation}
\lambda ^{(i)}\approx \frac{1}{\sum \left\langle \sigma _{ij}\right\rangle
n_{j}},  \label{mean-free}
\end{equation}
so $\lambda _{scat}^{(i)}$ represents the lower limit for $\tau
_{scat}^{(i)} $. For example, the pion mean free path in the rest frame of the fluid element at freeze-out, $\lambda ^{(\pi )}(\tau _{f.o.})$, was  roughly estimated as follows

\begin{equation}
\lambda ^{(\pi )}(\tau _{f.o.})\approx \frac{1}{\sigma _{\pi p}n_{B}+\sigma
_{\pi \pi }n_{M}}  \label{mean-ap}
\end{equation}
with the parameters
 $\sigma _{\pi p}=65$ mb  to be total cross section for
pion-proton scattering, and  the same cross-section for all
non-strange baryons,\ whereas $\sigma _{\pi \pi }=10$ mb is the total cross-section for pion-pion scattering and the same cross-section for all non-strange mesons. 

In the above approximation for fluid decoupling, the following equations should be satisfied
\begin{equation} 
\tau_{scat}(T(x),\mu(x))=\tau_{exp}(T(x),\mu(x))
\label{tau-scatt}
\end{equation}
where $T(x)$ and $\mu_B(x)$ are the local temperature and  baryon chemical potential correspondingly.
It is a complicated but promising way to build the true-like decay hypersurface for near-local equilibrium {\it baryon-reach} expanding matter. 
	
Another criterion that is based on the energy  densities in the fluid can be
\begin{equation}
\epsilon(T(x),\mu_B(x)) = \epsilon_{dec} = const. 
\label{eps}
\end{equation}
 However, there is only hope that there exists some parameter  $\epsilon_{dec}$ when the criteria (\ref{tau-scatt}) and (\ref{eps}) will nearly coincide. In this paper, we will follow simpler criteria of decoupling based on the energy density (\ref{eps}).
  
The realistic situation, however, can be more complicated, even at the selected criteria: in the case of low collision energies, the system may never reach complete thermalization, but only if possible, a partial one. In such a case, despite near full thermalization at the particlization stage at high enough energies, the system at the decaying (into the interacting particles) stage consists of two expanding components of baryon-rich matter: particle gas and fluid. The most natural way then is to select the energy density interval when the mechanism, forming the dense (QGP?) droplets, stops working, and soon after the corresponding (see above) hydrodynamical part of the system decays into hadrons without reaching full thermalization. In our approximation, it means that one has to transform the hydrodynamically involved part of the system into particles at this droplet's (mean) decay time.  So, because we do not know the quantum component of the QCD fluctuations, the corresponding energy density at which the decay of the hydro-component at small relativistic energies happens is a free parameter. The corresponding particle injection from this decay of the hydro-component in the case of not full thermalization is just added to the preserved yet  UrQMD component. Of course, the local energy-momentum conservation law for the evolution and hydro-decays of the total system, consisting of both UrQMD (or SMASH) and hydro-components, must be implemented in any scenario. It is done in the iHKM{\it e} (like in iHKM) scenario that is developed in the presented article.

However, as we already mentioned, switching from quasiclassical microscopic models (UrQMD, SMASH, etc) to the macroscopic hydrodynamic regime could be impossible if one naively tries to base it on a distribution function from a single transport simulation,  as it brings significant fluctuations in coordinates and anisotropies in momentum space, contradicting the basic assumptions of hydrodynamics. So, one needs to use some averaging procedure to provide smooth initial conditions for a subsequent description of hydrodynamic expansion and, of course, especially, for event-by-event analysis.

There are two common solutions to this problem. The straightforward one is to generate huge amounts of events and then average over them. However, to study the influence of fluctuations in the initial conditions on the final observables, one needs to introduce some similarity criteria between events. Of course, averaging over the centrality class might be too rough and unsuitable for event-by-event analysis. The common approach to this problem is applying some Gaussian smearing procedure to each particle and then constructing only the time component of the stress-energy tensor $T^{0\mu}$ and baryon current $J^{0}$. The other components are restored using an equation of state and explicit representation of the relativistic hydrodynamic tensors through macroscopic fields of velocity $u^{\mu}$, energy-density $\epsilon$, pressure $p$, and baryon density $n_B$. 

In the paper~\cite{Oliinychenko:2015lva}, a comparison of such a procedure across several hybrid models is presented. Additionally, the same paper introduces a Lorentz-invariant Gaussian kernel for particle smearing over the space but at constant time $t$ (See also \cite{Petersen:2008dd})
\begin{equation}\label{eq:kernelOliinychenko}
    {\cal K}({\bf r}) = \frac{\gamma}{(2\pi\sigma^2)^{3/2}} \exp\left( \frac{-{\bf r}^2 - ({\bf r\cdot u})^2}{2\sigma^2}\right),
\end{equation}
where ${\bf u}$ is the velocity of the particle, ${\bf r}$ is the vector from the particle's position to the spatial point where its contribution is evaluated, and $\gamma$ is the Lorentz contraction factor. Such a procedure can be attributed to the averaging over an ensemble of ``similar'' collision events without generating them. The similarity between the events is described by the $\sigma$ parameter. 

In this paper, we propose a modification of the kernel in a covariant form that enables Gaussian smearing in Milne coordinates, which are more suitable for our model (see Appendix~\ref{sec:appendixA} for details). For a particle \( i \) with baryon charge \( B_i \), mass \( m_i \), momentum \( p^{\mu}_i \), velocity \( u_i^{\mu} = p_i^{\mu}/m_i \), and spatial position \( x_i^{\mu} \), the relative contribution of its energy, momentum, and charge to the lattice grid cell \( \Delta \sigma^{\mu}_j \), centered at \( x_j^{\mu} \), is given by

\begin{equation}\label{eq:kernelIHKM}
{\cal K}_{ij}= \frac{n_{i}^{\lambda}  u_{i
\lambda} }{(\pi R^2)^{3/2}} \exp\left( \frac{ r_{ij}^{\mu} \left( g_{\mu\nu} - u^{i}_{\mu}u^{i}_{\nu} \right) r_{ij}^{\nu} }{R^2}\right).
\end{equation}

Here, $r_{ij}^{\mu} = x^{\mu}_{i}-x^{\mu}_j$ represents the radius vector between the particle and the center of the cell $\left( x_i, x_j \in \sigma^{\mu}\right)$, $n_i^{\mu}$ is the normal vector to hypersurface at point $x_i$, while $R$ is a free scalar parameter.

Many hydrodynamic models~\cite{Schenke:2010nt, Pang:2018zzo, DelZanna:2013eua, Bazow:2016yra}, including the vHLLE code~\cite{Karpenko:2013wva} employed in both the previous and current versions of the iHKM model, operate in Milne coordinates. These coordinates, defined by the proper time \(\tau = \sqrt{t^2 - z^2}\) and spacetime rapidity \(\eta = \tanh^{-1}(z/t)\), are particularly suitable for high-energy collisions, where an approximately boost-invariant longitudinal expansion is expected. The use of hyperbolic (Milne) coordinates also allows one to fix the number of grid points in the longitudinal direction $N_{\eta}$ in advance, by scaling the grid spacing with proper time $\tau$  \cite{Cimerman:2023hjw} instead of rescaling the full longitudinal extent of the system ($z_{\text{max}}$ grows almost with the speed of light), as would be necessary in Cartesian coordinates. This feature is especially useful for models that aim to describe a wide range of collision energies within a unified framework, as we do in iHKM.

As we use Milne coordinates for the system's evolution in the subsequent stages, it is natural to choose a hypersurface of constant proper time \(\tau\) in the kernel~(\ref{eq:kernelIHKM}). For a fluid cell located at space-time rapidity \(\eta\), the hypersurface element is given by
\begin{align}\label{eq:milneNormal}
    \Delta \sigma^{\mu} &= n^{\mu} \, \Delta x \, \Delta y \, \Delta \eta, \\
    n^{\mu} &= \left( \tau \cosh \eta,\, 0,\, 0,\, \tau \sinh \eta \right),
\end{align}
where \(\Delta x\), \(\Delta y\), and \(\Delta \eta\) are the cell sizes in the transverse and longitudinal directions, with numerical values of \(0.3\,\mathrm{fm}\), \(0.3\,\mathrm{fm}\), and \(0.05\), respectively.

Utilizing kernel~(\ref{eq:kernelIHKM}), we obtain the following inputs for the next stages of the model:
\begin{align}\label{eq:urqmd_tesors}
    &T^{\mu\nu}_{\text{urqmd}} (\tau; x_j) = \sum_{i}\frac{p_i^{\mu}p_i^{\nu}}{p_i^{0}} {\cal K}_{ij},\\
    &J^{\mu}_{\text{urqmd}}(\tau; x_j)  = \sum_{i} B_i \frac{p_i^{\mu}}{p_i^{0}} {\cal K}_{ij},
\end{align}
where the sums are taken over all the particles from UrQMD evolution that satisfy the condition 
\begin{equation}\label{eq:timeselection}
    \left|\sqrt{t_i^2-z_i^2} - \tau \right| < \Delta \tau/2\,.
\end{equation}

Notably, we have not found any publications that apply particle smearing directly on a hypersurface of constant proper time \(\tau\). Instead, such procedures are typically performed at constant laboratory time \(t\), followed by a transformation to Milne coordinates. While we do not undertake a detailed comparison of these two approaches, we emphasize that the kernel~(\ref{eq:kernelIHKM}) retains a Lorentz-covariant form, ensures particle number conservation, and accounts for Lorentz contraction along particle velocities (for more details, see Appendix~\ref{sec:appendixA}). Moreover, it enables an efficient reconstruction of the non-equilibrium distribution function on the particlization hypersurface (see Section~\ref{sec:particlization}) by storing only 400 to 1000 particles (depending on the collision energy \(\sqrt{s_{NN}}\)) whose trajectories satisfy the selection criteria specified in Eq.~(\ref{eq:timeselection}) \footnote{With more common kernel~\ref{eq:kernelOliinychenko} one needs to iterate over all tracks between $t_{\text{min}}=\tau$ and $t_{\text{max}} = \sqrt{\tau^2 + z_{\text{max}}^2}$, where $z_{\text{max}}$ is longitudinal coordinate of leading particles at proper time $\tau$.}.

Lastly, we note that in the results presented in this paper, for simplicity, we do not consider separate conservation equations for electric charge and strangeness. Instead, we assume local constraints \(n_s = 0\) and \(n_q = A/Z \cdot n_B \approx 0.4\,n_B\), appropriate for gold nuclei. Although both UrQMD and vHLLE support such calculations (but not the equations of state considered in this paper), we omit them here to reduce computational complexity and focus on the primary features of the model.

\subsection{Thermalization}
The thermalization stage is one of the distinctive features of iHKM. During this stage, the matter can be phenomenologically decomposed into two distinct components: a (near) locally equilibrated component, described by macroscopic fields, and a non-thermalized component, represented by out-of-equilibrium hadrons and strings evolving via UrQMD in the cascade regime. Both components contribute to the non-equilibrium energy-momentum tensor~\cite{Naboka:2014eha} and the charge currents, ensuring that the corresponding equations respect the conservation laws for the total system.
This stage is the primary difference between our model and more common models with instant thermalization
\begin{equation}\label{tensorSplit1}
    T^{\mu\nu}_{\text{total}}(x) = T^{\mu\nu}_{\text{urqmd}}(x) \cdot {\cal P}_{\tau} + T^{\mu\nu}_{\text{hydro}}(x)\cdot \left( 1- {\cal P}_{\tau} \right),
\end{equation}
\begin{equation}\label{tensorSplit2}
    J^{\mu}_{\text{total}}(x) = J^{\mu}_{\text{urqmd}}(x) \cdot {\cal P}_{\tau}+ J^{\mu}_{\text{hydro}}(x)\cdot \left( 1- {\cal P}_{\tau} \right),
\end{equation}
where ${\cal P}_{\tau} = {\cal P}(\tau)$ is a weight function such that ${\cal P}(\tau_0)=1$ at the start of the thermalization stage, ${\cal P}(\tau_{th})=0$ at the end, and $0<{\cal P}(\tau_0<\tau<\tau_{th})<1$ in between. Its explicit form will be discussed later. Both the total stress-energy tensor and the baryon current obey conservation laws:
\begin{equation}
\partial_{\mu}T^{\mu\nu}_{\text{total}}(x)= 0; \quad  \partial_{\mu} J_{\text{total}}^{\mu}(x) = 0.
\end{equation}
Exploiting the conservation laws accounted for in UrQMD evolution 
\begin{equation}
    \partial_{\mu}T^{\mu\nu}_{\text{urqmd}}(x)=0, \quad \partial_{\mu}J_{\text{urqmd}}^{\mu}(x)=0\,,
\end{equation}
we obtain hydrodynamic-like equations with a source for rescaled tensors
\begin{equation}\label{eq_for_T}
    \partial_{\mu} \Tilde{T}^{\mu\nu}_{\text{hydro}}(x) = - T^{\mu\nu}_{\text{urqmd}}(x) \cdot \partial_{\mu}  {\cal P}_{\tau},
\end{equation}
\begin{equation}\label{eq_for_J}
   \partial_{\mu}\Tilde{J}^{\mu}_{\text{hydro}}(x)     = - J^{\mu}_{\text{urqmd}}(x) \cdot \partial_{\mu}{\cal P}_{\tau}.
\end{equation}
Here, the re-scaled (tilded) hydrodynamic tensors are defined as
\begin{equation}\label{rescT}
    \Tilde{T}^{\mu\nu}_{\text{hydro}}(x) = T^{\mu\nu}_{\text{hydro}}(x)\cdot \left( 1 - {\cal P}_{\tau}\right) ,
\end{equation}
\begin{equation}\label{rescJ}
    \Tilde{J}^{\mu}_{\text{hydro}}(x) = J^{\mu}_{\text{hydro}}(x)\cdot \left( 1 - {\cal P}_{\tau}\right),
\end{equation}
As in the previous papers \cite{Akkelin:2009nz, Naboka:2014eha}, we utilize the same ansatz inspired by the Boltzmann equation in relaxation time approximation, with probability  ${\cal P}(\tau)$
\begin{equation}\label{eq:Ptau}
    {\cal P}(\tau ) =  \left(\frac{\tau_{th}-\tau}{\tau_{th}-\tau_0} \right)^{\frac{\tau_{th}-\tau_{0}}{\tau_{rel}}}.
\end{equation}
Wherein a free parameter of the model $0 < \tau_{rel} < \tau_{th} - \tau_0$ is introduced. The relaxation time $\tau_{rel}$ characterizes the rate of the thermalization process. To avoid introducing additional freedom during the model calibration, in this paper, we set the relaxation time at its minimum value $\tau_{rel} = \tau_{th} - \tau_{0}$. At the same time, the influence of non-thermal dynamics can be varied via the thermalization time $\tau_{th}$.

Solving Eqs. (\ref{eq_for_T}) and (\ref{eq_for_J}) constitutes the primary objective of the relaxation stage in the model. These equations are employed to update, over time $\tau$, the values of the time components of the corresponding tensors $\Tilde{T}^{0\mu}_{\text{hydro}}$ and $\Tilde{J}^{0}_{\text{hydro}}$ for each cell of the spatial grid. The remaining components are restored using the Israel-Stewart form \cite{Israel:1979wp} of the tensors:
\begin{equation}\label{Ttilde}
    \frac{\Tilde{T}_{\text{hydro}}^{\mu\nu}}{1-{\cal P}_{\tau}} = T^{\mu\nu}_{\text{hydro}}(x) = \left( \epsilon + p \right) u^{\mu}u^{\nu} - pg^{\mu\nu} + \pi^{\mu\nu}.
\end{equation}
\begin{equation}\label{eq:Jtilde}
    \frac{\Tilde{J}_{\text{hydro}}^{\mu}}{1-{\cal P}_{\tau}} = J^{\mu}_{\text{hydro}}(x) = n_{B} u^{\mu},
\end{equation}

where $\epsilon$, $p$, and $n_B$ represent the local energy density, pressure, and baryonic density, respectively. $u^{\mu}$ denotes the four-velocity of the fluid, $g^{\mu\nu}$ is the metric tensor, and $\pi^{\mu\nu}$ stands for the shear-stress tensor. The local energy density, pressure, and four-velocity are derived from $T^{0\mu}_{\text{hydro}}$ utilizing the equation of state $p=p(\epsilon, n_B)$. At the same time, the shear-stress tensor evolves according to an independent equation within the Israel-Stewart framework \cite{Israel:1979wp, Naboka:2015qra}. To numerically solve Eqs.~(\ref{eq_for_T}) and (\ref{eq_for_J}), we utilize the vHLLE code \cite{Karpenko:2013wva} with modifications adjusting source terms \cite{Naboka:2015qra}. We do not consider other transport coefficients, such as bulk pressure, diffusion, and heat conductivity, in this paper.  We employ two different equations of state, namely the chiral EOS \cite{Steinheimer:2010ib} with a crossover-type transition between QGP and the hadron stages and an EoS with the first-order phase transition proposed in \cite{Kolb:2003dz}.

\subsection{Hydrodynamic expansion}

We suppose, that not only at ultrarelativistic heavy ion collisions but also, at least, at intermediate energies at BES RHIC, at some $\tau=\tau_{th}$ the system attains a state of local near-equilibrium, characterized by hydrodynamic tensors $T^{\mu\nu}_{\text{total}}=T^{\mu\nu}_{\text{hydro}}$ and $J^{\mu}_{\text{total}}=J^{\mu}_{\text{hydro}}$, with the source terms in Eqs.~(\ref{rescT}) and (\ref{rescJ}) vanishing. The description of ultrarelativistic (top RHIC and LHC energy within iHKM \cite{Naboka:2015qra} is naturally used just this approach. Below, we will discuss the more complicated situation. In simple cases, the hydrodynamic evolution persists until the system becomes dilute and departs from partial local equilibrium, which means that the hydrodynamic approximation breaks down, and particle language is necessary. Then the system is appropriately described in microscopic terms using a hadron-resonance gas model, UrQMD, in the case of this study. 

The anticipated outcome of the hydrodynamic stage is the formation of a hypersurface that marks the transition between the fluid and gas phases, often referred to as the 'particlization hypersurface', and denoted as $\sigma_{\text{sw}}$ in this paper. As we wrote before,  we carry out this transition at a fixed energy density $\epsilon_{\text{sw}}$ for simplicity. In our model, the construction of the particlization hypersurface is achieved using the Cornelius routine \cite{Huovinen:2012is, Molnar:2014fva}.

When dealing with low-energy collisions at $\sqrt{s_{NN}}\sim 3-10$ GeV, as observed in experiments such as HADES, RHIC BES, or CBM, it is reasonable to anticipate that the system may not achieve {\it complete} thermalization. Consequently, extending the particlization hypersurface criteria $\sigma_{\text{sw}}$ into the thermalization phase, $\tau < \tau_{th}$, becomes imperative.

During the thermalization period, it is noteworthy that there exist formally three distinct ways to define energy density for total particlization
\begin{enumerate}
    \item The local equilibrium energy density $\epsilon_{\text{hydro}}$, derived from $T_{\text{hydro}}^{\mu\nu}$.
    \item The non-equilibrium energy density $\epsilon_{\text{urmqd}}$ derived from $T_{\text{urqmd}}^{\mu\nu}$.
    \item The mixed energy density $\epsilon_{\text{total}}$ derived from the total energy-momentum tensor $T_{\text{total}}^{\mu\nu}$.
\end{enumerate}

While it is conventional to associate the particlization hypersurface with the total energy density, smoothed out along the system, we opt to utilize $\epsilon_{\text{hydro}}$ instead when hydrodynamic fluid fills out the essential (central) part of the system. If the matter may not be fully thermalized at quite small collision energies, $\cal{P}$ becomes a free parameter instead of $\tau_{th}$. 
We found in preliminary calculations that such an approach is not bad for the energies $\sqrt{s_{NN}}>8$~GeV.  So, at such energies, it is possible that hydrodynamics inject essentially the thermal particles into expanding (and still existing!) UrQMD-system. Then the totally non-locally equilibrated system will include initially (locally)  thermal particles from hydrodynamics. The latter can be easily calculated by using a generalized Cooper-Frye prescription with collective velocities $u^{\mu}$ of the pure hydrodynamical part accounting for a known equation of state. It is notably simpler and faster than the utilization for building the decay hypersurface for the total momentum-energy tensor. Notice, in addition, that the criteria described in subsection~\ref{subsec:IIA} apply to locally equilibrated systems but not to mixtures of equilibrated and non-equilibrated systems. The proposed method addresses several technical challenges simultaneously.

The transport model UrQMD employs an EoS corresponding to a hadron resonance gas, whereas the hydrodynamic stage may use a different EoS, such as one incorporating a QCD crossover or a phase transition. This mismatch in EoS can lead to so-called matching artifacts, including discontinuities in pressure, entropy non-conservation, and spurious flows at the transition hypersurfaces. While energy and momentum conservation can be ensured during the mapping procedures, the differences in thermodynamic assumptions between models can introduce systematic effects in the final-state observables. These artifacts are a known limitation of hybrid approaches and should be kept in mind when interpreting results.

For very low relativistic energies (near 2 GeV per nucleon pair) there can be only a small part of the system might be involved in hydrodynamic motion, and complicated criteria (as in item 3 above)  for transition of the total system into hadron gas have to be used.

A few technical details are worth noting. Minor time fluctuations can appear in the UrQMD energy-momentum tensor components due to the transformation from Cartesian to Milne coordinates during the thermalization stage, as described in Eq.~(\ref{eq:timeselection}) and discussed further in Appendix~\ref{sec:appendixA}. These fluctuations vanish when averaging over several time steps. As a result, the transition from the particle-based source to the hydrodynamic energy-momentum tensor, governed by Eq.~(\ref{eq_for_T}), becomes smooth. Empirically, we find that averaging over eight consecutive time steps is sufficient to ensure that the hydrodynamic component evolves smoothly in time, ultimately leading to a well-behaved particlization hypersurface.

It is also important to note that at the initial proper time when the construction of the particlization hypersurface begins, some regions of the system already have an energy density below the particlization threshold \( \epsilon_{\text{sw}} \). These regions are typically located at the system's periphery (the so-called corona) or in the spectator zones. Particles in these areas generally do not undergo thermalization and moreover, are ignored by the Cornelius module.

To ensure the conservation of total energy, momentum, and conserved charges in the system, we manually extend the particlization hypersurface to include these non-thermalized regions. For such extensions, we assign hypersurface elements a normal vector corresponding to constant proper time, as given in Eq.~(\ref{eq:milneNormal}), and set \( \mathcal{P}_{\tau} = 1 \) for them—indicating that these regions are purely non-equilibrated.

\subsection{Particlization}\label{sec:particlization}
Following the transition through the particlization hypersurface, the matter becomes sufficiently dilute to be effectively described as a hadron-resonance gas. In the current version of iHKM, hadrons are injected from two distinct sources: the equilibrium (eq) and non-equilibrium (n.eq) components. Analogous to the decomposition in Eqs.~(\ref{tensorSplit1},~\ref{tensorSplit2}), the system’s distribution function is expressed as the sum of these two contributions:
\begin{equation}
    f(x,p) = f_{\text{n.eq}}(x,p) \cdot \mathcal{P}_{\tau} + f_{\text{eq}}(x,p) \cdot \left(1 - \mathcal{P}_{\tau}\right)\,,
\end{equation}
where \( f_{\text{n.eq}}(x,p) \) corresponds to non-equilibrated hadrons originating from the same UrQMD event used for the initial dynamics and is constructed using the kernel~(\ref{eq:kernelIHKM}), while \( f_{\text{eq}}(x,p) \) represents the standard locally near-equilibrium component typically used in hydrodynamic-to-transport switching procedures. Notably, once the thermalization time is reached, \( \mathcal{P}_{\tau} \to 0 \), and all particles are emitted exclusively from the near-equilibrium source, consistent with conventional hybrid models.

Both components of the distribution function can be expressed as a sum over all hadron species \(i\):
\begin{equation}
    f_{\text{eq}/\text{n.eq}}(x,p) = \sum_i f^{\text{eq}/\text{n.eq}}_i(x,p)\,,
\end{equation}
where, for the equilibrium component, the sum runs over all hadron species in the ideal hadron-resonance gas into which the hydrodynamic fluid is converted. For the non-equilibrium component, the sum includes all hadron species present in the UrQMD event used to generate the initial conditions. 

For each infinitesimal element of the particlization hypersurface \( \Delta \sigma^{\mu}(x) \), particles are then generated using the well-established Cooper–Frye prescription~\cite{Cooper:1974mv}, with minor modifications discussed later:
\begin{equation}\label{CooperFrye}
    N_i(x,p) = \frac{ \Delta \sigma_{\mu} p^{\mu} }{p^0} f_i(x, p)\,.
\end{equation}

Particles from both the equilibrium and non-equilibrium components are sampled independently using their respective distribution functions, \( f_{\text{eq}} \) and \( f_{\text{n.eq}} \), without applying the factors \( \mathcal{P}_{\tau} \) and \( 1 - \mathcal{P}_{\tau} \) directly during sampling. These weights are instead used to determine the relative fraction of particles from each component that are subsequently passed into the UrQMD afterburner stage.

\subsubsection{Emission of thermal particles}

The near-equilibrium distribution function \( f^{\text{eq}}_i(x, p) \) for each particle species depends on the local thermodynamic properties of the system: temperature, chemical potentials, and the shear-stress tensor. The fluid velocity \( u^{\mu}(x) \), energy density $\epsilon$, and baryon charge density \( n_B(x) \) are extracted from the hydrodynamic energy-momentum tensor, the net baryon current, and the equation of state using the standard Landau matching procedure:
\begin{equation}\label{eq:matching}
\begin{aligned}
    T^{0\mu}_{\text{hydro}} &= (\epsilon + p)\, u^0 u^\mu - p\, g^{0\mu}, \\
    J_{\text{hydro}}^0 &= n_B u^0, \\
    p &= p(\epsilon, n_B)\,.
\end{aligned}
\end{equation}

Note that these tensors are obtained from the tilded ones—derived from the hydrodynamic equations with source terms, Eqs.~(\ref{eq_for_T})--(\ref{eq_for_J})—by dividing them by the factor \( 1 - {\cal P}_{\tau} \), as shown in Eqs.~(\ref{Ttilde}) and~(\ref{eq:Jtilde}). Also, the components \( \pi^{0\mu} \) vanish here because the shear-stress tensor is orthogonal to the fluid four-velocity in the Landau frame~\cite{Israel:1979wp}.

However, when transitioning from the hydrodynamic description to the hadron gas phase, the corresponding equations of state (EoS) typically do not match exactly. This mismatch results in discontinuities in the thermodynamic properties of the system across the particlization hypersurface. In our model, we adopt the hadron gas EoS to solve Eqs.~(\ref{eq:matching}) and use the corresponding energy density as the switching criterion within the Cornelius routine. This approach ensures local—and therefore global—conservation of all conserved charges. However, it introduces small discontinuities: typically less than 1\% in \(\epsilon\) and \(n_B\), and around 10\% in \(p\).\footnote{In a statistically insignificant number of cases, these discontinuities may be much larger due to failures of the numerical algorithms.}

On the other hand, closeness of intensive quantities leads to noticeable discontinuities in extensive properties such as temperature \(T\) and baryon chemical potential \(\mu_B\). This situation can be improved by employing an equation of state for the hydrodynamic stage that smoothly connects to the hadron resonance gas properties across a wide range of the QCD phase diagram. In future work, we plan to implement one such EoS (NEOS BQS \cite{Monnai:2019hkn}) into our model.

When the local temperature and the baryon, electric, and strange chemical potentials are determined from the local energy density \( \epsilon \) and the conserved charge densities \( n_B \), \( n_Q = 0.4\, n_B \), and \( n_S = 0 \) using the equation of state, one can construct the equilibrium distribution function for the ideal hadron resonance gas. The chemical potential \( \mu_i \) for each particle species is then given by:
\begin{equation}
\mu_i = B_i\mu_B + q_i \mu_q + s_i \mu_s,
\end{equation}

where $B_i$, $q_i$, and $s_i$ represent the particle's baryon, electric, and strange charges, respectively. The near-equilibrium corrections to the distribution functions are obtained after applying the Grad ansatz \cite{Grad:1949zza} for viscous corrections. Assuming the same corrections for all hadron species, the thermal particle production in the rest frame of the fluid can be written as:
\begin{equation}\label{CFangle}
\begin{aligned}
\frac{d^3N_{i}}{dp^{}d(\cos \theta) d\phi} &= \left( 1- {\cal P}_{\tau}\right) \frac{d\sigma^{}_{\mu}p^{\mu}}{p_0^{}}p^{2} f_{\text{eq}} (p^{0}, T, \mu_i) \\
&\quad \times \left( 1+ (1\mp f_{\text{eq}})\frac{p^{}_{\mu} p^{}_{\nu}\pi^{\mu\nu}}{2T^2(\epsilon+p)}\right),
\end{aligned}
\end{equation}
Here, $\mp$ indicates Fermi/Bose statistics. For more detailed information, we refer the reader to the papers \cite{Naboka:2014eha} or \cite{Karpenko:2015xea}.

\subsubsection{Non-thermal emission}

As mentioned above, for the non-equilibrium distribution function, we use the same set of particles from the sub-ensemble constructed during the pre-equilibrium dynamics stage. Then, for each element of the particlization hypersurface \( d\sigma_j^{\mu} \) located at position \( x_j^{\mu} \), the probability that a particle with space-time coordinate \( x_i^{\mu} \) will be emitted is given by the product:

\begin{equation}
\begin{aligned}
    {\cal W}(p_i, x_i; d\sigma_j) &= \theta\bigl(p_i^{\mu} d\sigma^j_{\mu}\bigr) \times \theta\left(\frac{\Delta \tau}{2} - \left| \tau_j - \tau_i \right| \right) \\
    &\quad \times {\cal K}_{ij} \times \frac{p_i^{\mu} d\sigma^j_{\mu}}{p_i^0}.
\end{aligned}
\end{equation}

The first Heaviside step function ensures that particles move from the hotter to the colder phase. The next two factors arise from the construction of the non-equilibrium distribution function in Eqs.~(\ref{eq:kernelIHKM}) and~(\ref{eq:timeselection}), determining whether the particle intersects the hypersurface in space and time. The last term accounts for the size and space-time orientation of the hypersurface element. Summing over the entire hypersurface and all particle trajectories from the initial UrQMD event, weighted by \({\cal P}_{\tau}\), yields the total contribution of non-thermal emission:

\begin{equation}
    N_{\text{n.eq}} = \sum_{i \in \{\text{tracks}\}} \sum_{j \in \{\sigma_{\text{sw}}\}} {\cal P}_{\tau_j} \, {\cal W}(p_i, x_i; d\sigma_j).
\end{equation}

\subsubsection{Non-space-like surface emission treatment}

It is well-known that the hypersurface of constant energy density might include problematic regions \cite{Sinyukov:1988vj} 'sink' terms with $d\sigma_0 < 0$ and non-space-like parts, with $d\sigma^{\mu}d\sigma_{\mu} < 0$. In both cases, not all particles near the surface can cross it $p^{\mu}d\sigma_{\mu} < 0$, leading to negative contributions in the Cooper-Frye formula (\ref{CooperFrye}).

To address this problem adequately, we adopt a prescription proposed in \cite{Sinyukov:1988vj, Amelin:2006qe}, which suggests substituting $p^{\mu}$ with a generalized momentum $\pi^{\mu}$ in the near-equilibrium distribution function in Eq.~(\ref{CFangle}).

\begin{equation}\label{generalizedMomentum}
\begin{aligned}
\pi^{\mu}(x,p) &= p^{\mu}\theta\left( 1 - \lambda \right) 
+ u^{\mu}\left(p\cdot u\right)\theta\left(\lambda - 1\right),
\end{aligned}
\end{equation}
where $\theta$ is the Heaviside step function and $\lambda$ is defined by:

\begin{equation}
\lambda = \lambda(x,p) = \left| 1- \frac{p\cdot n}{ \left(p\cdot u\right) \left(n\cdot u\right) } \right|.
\end{equation}

In this formula, $n^{\mu}$ represents the normal vector to $d\sigma$. This substitution modifies the distribution function in a manner that preserves the number of emitted particles but slightly violates energy conservation. We refer the reader to the paper \cite{Amelin:2006qe} for a more detailed explanation.

\subsubsection{Fluidization and Particlization Rates}

\begin{figure}[h!]
    \centering
    \includegraphics[width=0.47\textwidth]{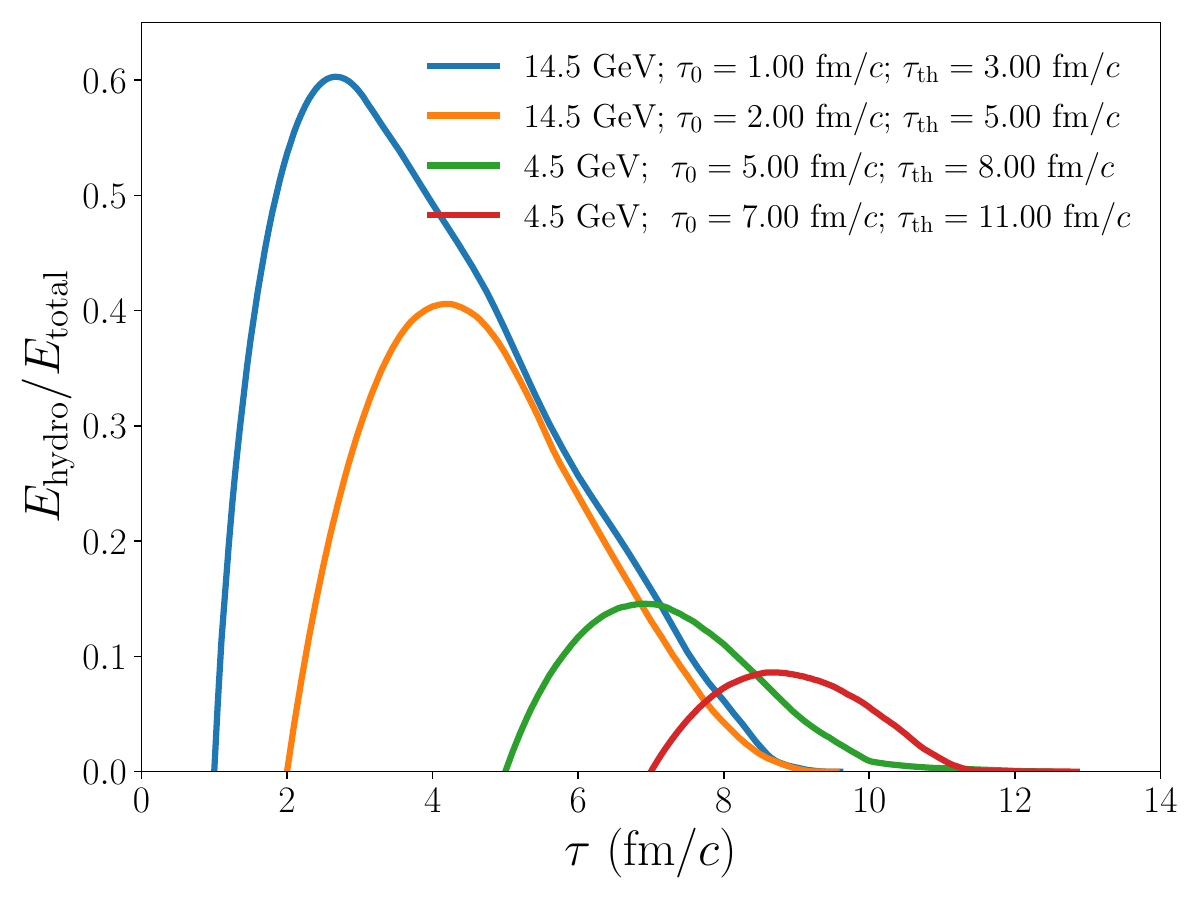}
    \caption{Ratio of the energy stored in the hydrodynamic part of the system to the total energy as a function of \(\tau\). For each of the lines, one event from $0-5$~\% centrality is considered.}
    \label{fig:efraction}
\end{figure}

Having discussed the mechanisms of converting particles to fluid (thermalization or fluidization) and the reverse process (particlization), we now turn to examining the rates of these processes in typical simulations. To describe these qualitatively, we calculate the ratio of the energy stored in the hydrodynamic part of the system—corresponding to the stress-energy tensor given by Eq.~(\ref{rescT}) (including the factor \(\mathcal{P}_{\tau}\))—to the total energy of the system, Eq.~(\ref{tensorSplit1}). The energy (in the laboratory frame) is computed as the flux of the \(T^{0\mu}\) component of the stress-energy tensor through the hypersurface of constant proper time. 

\begin{equation}
    E(\tau) = \int_{\tau} d\sigma_{\mu} \, T^{0\mu}\,,
\end{equation}

where for the hydrodynamic component we additionally exclude parts where $\epsilon<\epsilon_{\text{sw}}$ as they are needed for energy conservation and smooth boundary conditions for hydrodynamics, but they correspond to already hadronized matter. The ratio \(E_{\text{hydro}} / E_{\text{total}}\) as a function of \(\tau\) is shown in Fig.~\ref{fig:efraction}. The general trend is as follows:

\begin{itemize}
    \item Before the onset of the thermalization stage ($\tau < \tau_0$), the entire system is out of equilibrium and is described by the UrQMD cascade. In this work, we neglect the slow, ``natural'' thermalization that occurs within UrQMD (see \cite{Bravina:1999kd}).
    
    \item During the relaxation stage ($\tau_0 < \tau < \tau_{\text{th}}$), the dense regions of the system begin to thermalize, leading to a gradual increase in the hydrodynamic component.
    
    \item At the periphery of the system, fluid elements begin to hadronize, reducing the overall hydrodynamic fraction. As a result, the maximum contribution of the hydrodynamic component typically occurs before $\tau_{\text{th}}$. Figure~\ref{fig:efraction} shows that the maximum fluid fraction depends on the details of the thermalization process—namely, the values of $\tau_0$ and $\tau_{\text{th}}$—as well as on the collision energy $\sqrt{s_{NN}}$. Earlier thermalization and higher collision energies lead to a larger portion of the system reaching local equilibrium.
    
    \item After the thermalization time $\tau_{\text{th}}$—which may not be reached at very low collision energies or in non-central collisions—the system continues to expand, cool, and hadronize until the maximum energy density drops below the switching threshold $\epsilon_{\text{sw}}$.
    
    \item Beyond this point, the entire system is described as a hadron gas.
\end{itemize}

In this context, let us briefly discuss the similarities and differences between our approach and other hybrid models. Similar to our model, the frameworks presented in~\cite{Schafer:2021csj} and~\cite{Karpenko:2015xea} employ transport models (SMASH and UrQMD, respectively) and operate in Milne coordinates. However, unlike our approach, these models do not include an explicit thermalization stage. Instead, they assume instantaneous thermalization, typically occurring at the time of nuclear overlap (see Eq.~(\ref{eq:overlap}) in the next chapter). In our model we expect hierarchy of times $\tau_0 < \tau_{\text{overlap}} < \tau_{\text{th}}$. The differences between the models become increasingly significant at lower collision energies, particularly below 5~GeV, where we expect only partial thermalization of the system. Moreover, dissipative effects during the early stages are expected to be more pronounced in our model due to the non-equilibrium component, which may lead to different inferred values for transport coefficients.

Compared to the version of iHKM developed for ultrarelativistic energies~\cite{Naboka:2015qra}, which leads to full thermalization of the bulk of the system at some hypersurface, the current version additionally incorporates a smooth particlization of the system during this stage. At LHC energies, the matter density is significantly higher and thermalization occurs much earlier; as a result, the effects of gradual particlization were negligible due to their small magnitude.

In contrast to other models with continuous fluidization~\cite{Akamatsu:2018olk, Du:2018mpf, Shen:2017bsr}, we do not explicitly prescribe a specific mechanism for thermalization. As a result, our model lacks predictive power in certain aspects; however, it is more general and flexible. In principle, our approach allows for the possibility of constraining the thermalization mechanism by implementing more sophisticated forms of the function \({\cal P}\) (see Eq.~\ref{tensorSplit1}), for example, by allowing it to depend on local energy density. Such extensions, however, have not yet been implemented. 

Concerning the thermalization rate, Ref.~\cite{Akamatsu:2018olk} presents the evolution of the fluid fraction as a function of time at various collision energies, which enables a direct comparison with our model. It appears that at lower energies (below \(\sqrt{s_{NN}} \sim 10\)~GeV), the maximum fluid fraction is reached near the nuclei overlap time, while at higher energies it is attained later. This behavior is particularly important for the development of hydrodynamic flow and therefore significantly affects key observables such as transverse momentum spectra, flow harmonics, and the sizes of the particle-emitting source extracted via interferometric analysis. In particular, for \(\sqrt{s_{NN}} = 14.5\)~GeV—the energy used for demonstration in this work—the hydrodynamic phase in central collisions is expected to end around \(9\)~fm/\(c\), while in Ref.~\cite{Akamatsu:2018olk}, this duration can exceed \(15\)~fm/\(c\) in the central region of the system.

Another notable difference from other models concerns the treatment of core--corona separation, which is implemented in the aforementioned works as well as in other studies (e.g.,~\cite{Stefaniak:2022pxc}). In iHKM, the corona is also spatially and temporally separated at the periphery of the system, similar to those models. However, a non-thermal component is also present in the dense region of the system, in the sense of ensemble averaging.

\subsection{Hadronic cascade}

At the final post-hydrodynamical stage of the system's evolution, all particles from both equilibrium and non-equilibrium sources are input into the UrQMD hadron cascade code \cite{Bleicher:1999xi}. In the iHKM framework, we aim to account for all reliably known hadron resonance states, even those not processed by UrQMD. Therefore, heavy resonances not present in the UrQMD particle database are decayed right at $\sigma_{\text{sw}}$ to ensure energy-momentum conservation.

We generate 20 to 200 UrQMD events based on a single hydrodynamic run to increase statistics in event-by-event simulation. A detailed discussion of this procedure can be found in \cite{Holopainen:2010gz}. It's worth noting that this approach saves CPU time while, as demonstrated in similar models such as \cite{Karpenko:2015xea} and the iHKM analysis, it does not significantly impact the final observables, including Bose-Einstein correlations. However, we expect that artificial correlations might be present at several GeV energies when the multiplicities of thermal particles are relatively low -- due to this procedure. Therefore, it is reasonable to stick with a pure event-by-event simulation. However, this should not be the case for the $14.5$~GeV energies considered in all the simulation results presented in this paper.

\section{Model calibration and results}
For demonstration purposes in this paper, we focus exclusively on Au+Au collisions at $\sqrt{s_{NN}}=14.5$ GeV within the RHIC BES program. This energy serves as an intermediate point between lower-energy collisions at a few GeV and higher-energy collisions reaching several tens of GeV.

\subsection{Smoothing procedure}
To perform hydrodynamics simulations, starting with a smooth initial distribution of thermalized matter is necessary. However, capturing event-to-event fluctuations in the system's initial state is essential for accurately reproducing experimental data. Models that rely on a transport approach to describe pre-thermal dynamics typically derive the distribution function from a single event by applying Gaussian smearing to point-like particles using a kernel similar to Eq.~(\ref{eq:kernelOliinychenko}) or Eq.~(\ref{eq:kernelIHKM}), introducing free parameters into the model \cite{Schafer:2021csj, Cimerman:2023hjw, Oliinychenko:2022uvy, TMEP:2022xjg}. 

\begin{figure*}
    \centering
    \includegraphics[width=0.47\textwidth]{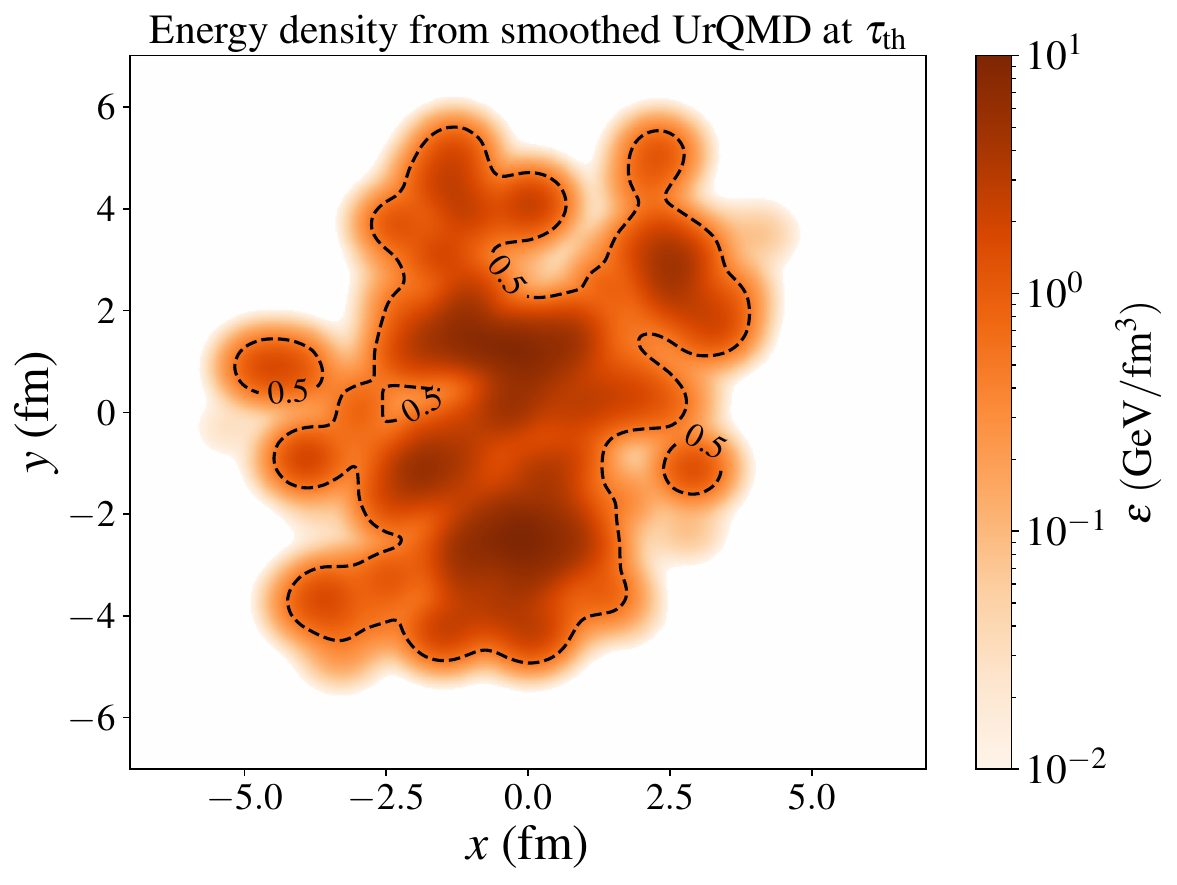}
    \includegraphics[width=0.47\textwidth]{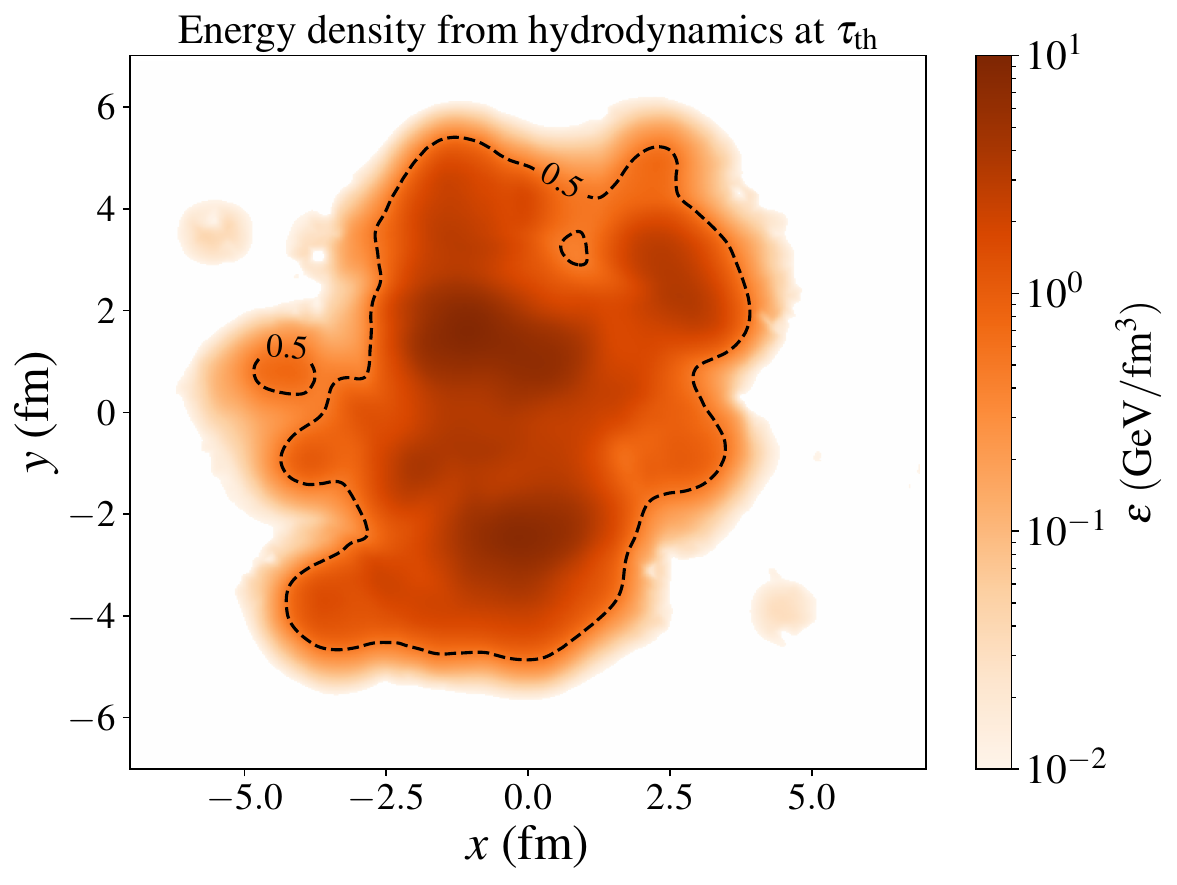}
    
\caption{Transverse-plane distribution of the energy density (logarithmic scale) at $z=0$ for a single Au+Au collision event at $\sqrt{s_{NN}}=14.5$~GeV in the 20–30\% centrality class. Left: energy density extracted from the UrQMD tensor~(\ref{eq:urqmd_tesors}), constructed using a Gaussian kernel~(\ref{eq:kernelIHKM}) with $R=0.5$~fm, at the thermalization proper time $\tau=\tau_{\rm th}=2.6$~fm/$c$. 
Right: energy density obtained from the hydrodynamic tensor in iHKM simulations at the same $\tau_{\rm th}$ (other model parameters chosen according to Set~1 in Table~\ref{tab:optimalParams}). Black dashed lines separate regions with energy density above and below $0.5$~GeV/fm$^3$.}

    \label{fig:energyDensity}
\end{figure*}

A distinctive feature of iHKM is the presence of a continuous thermalization stage, which further smooths the distribution of matter. In Figure~\ref{fig:energyDensity}, we illustrate the energy density distribution in the transverse plane at zero space-time rapidity $\eta$ for scenarios with and without the thermalization stage at the same proper time $\tau_{th}$. This noticeable difference can significantly influence the final observables and, consequently, our estimation of the model's optimal parameters, including the equation of state and transport coefficients.

\subsection{Free parameters and calibration}

Let us briefly summarize the free parameters of the model. They can be categorized into three groups:

\begin{enumerate}
    \item Responsible for the thermalization stage: 
        \begin{itemize}
            \item $\tau_0$ - start of thermalization stage
            \item $\tau_{rel}$ - relaxation time
            \item $\tau_{th}$ - end of thermalization stage
        \end{itemize}
    \item Smoothing parameter $R$
    \item Thermodynamical properties
        \begin{itemize}
            \item equation of state
            \item transport coefficients, e.g. $\eta/s$
            \item particlization energy density $\epsilon_{\text{sw}}$
        \end{itemize}
\end{enumerate}

To calibrate $\tau_0$ and $\tau_{th}$ we utilize experimental data for $\pi^{-}$ transverse momentum spectra varying values of these parameters around the typical scale of

\begin{equation}\label{eq:overlap}
    \tau_{\text{overlap}} = \frac{2R_N}{\sqrt{(\sqrt{s_{NN}}/2m_{N})^2 - 1}},
\end{equation}

It represents the time required for two nuclei to overlap completely as they move with their initial rapidities. In this equation, $R_N$ stands for the radius of one nucleus, and $m_{N}$ stands for the nucleon mass. To simplify the model calibration, reduce the number of free parameters, and save CPU time, we use a simple ansatz for the relaxation time, fixing it to a maximum allowed value as: $\tau_{rel} = \tau_{th} - \tau_{0}$. 

To determine the shear viscosity $\eta/s$, we start from the typical minimal value of $1/4\pi$ \cite{Kovtun:2004de} and increase it if the flow anisotropy $v_2$ is too strong in non-central collisions compared to the experimental data.

For the transition to the afterburner stage, we use a typical value for the switching energy density, $\epsilon_{\text{sw}} = 0.5$~GeV/fm$^3$, which lies in the range where the equations of state (EoS) for the liquid and gas phases of strongly interacting matter are close. However, reducing this value may improve the results, due to the less intense hadron annihilation processes in UrQMD compared to local-equilibrium hydrodynamics. In particular, we observe a noticeable sensitivity of the $\bar{p}/p$ ratio to $\epsilon_{\text{sw}}$, which motivates us to treat $\epsilon_{\text{sw}}$ as a free parameter when necessary. Similar sensitivity of this ratio to the details of the EoS and the particlization energy density has been reported in other studies (see, e.g.,~\cite{Monnai:2019hkn}), and has been addressed in numerous theoretical works~\cite{Garcia-Montero:2021haa, Becattini:2012sq}. The final sets of parameters used to simulate Au+Au collisions at $\sqrt{s_{NN}} = 14.5$~GeV with two different equations of state are summarized in Table~\ref{tab:optimalParams}.

\begin{table}[b]
\caption{\label{tab:optimalParams}
Parameters of iHKM providing the best description of bulk observables for Au+Au collisions at $\sqrt{s_{NN}}=14.5$~GeV.}
\begin{ruledtabular}
\begin{tabular}{c|c|c|c|c|c|c}
Title  & EoS & $R$    & $\tau_0$ & $\tau_{th}$ & $\eta/s$ & $\epsilon_{\text{sw}}$  \\
\hline
Set 1& Chiral\footnotemark[1]  & 0.5~fm & 1.2 fm/$c$ 
& 2.6 fm/$c$ & 0.08 & 0.5 GeV/fm$^3$\\
Set 2& PT1\footnotemark[2] & 0.5~fm & 1.4 fm/$c$ 
& 1.8 fm/$c$ & 0.08 & 0.35 GeV/fm$^3$ \\

\end{tabular}
\end{ruledtabular}

\footnotetext[1]{EoS with crossover transition from \cite{Steinheimer:2010ib}.}
\footnotetext[2]{EoS with first-order phase transition from \cite{Kolb:2003dz}.}

\end{table}

\subsection{Bulk observables}
\begin{figure*}
    \centering
    \includegraphics[width=0.47\textwidth]{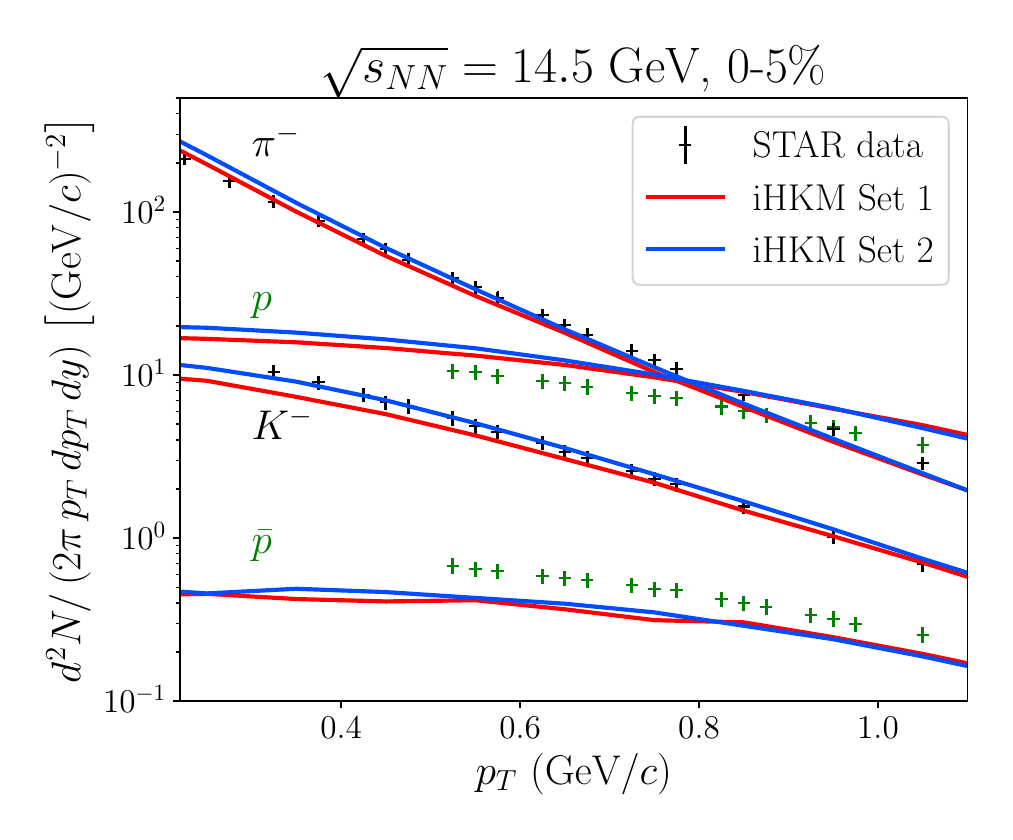}
    \includegraphics[width=0.47\textwidth]{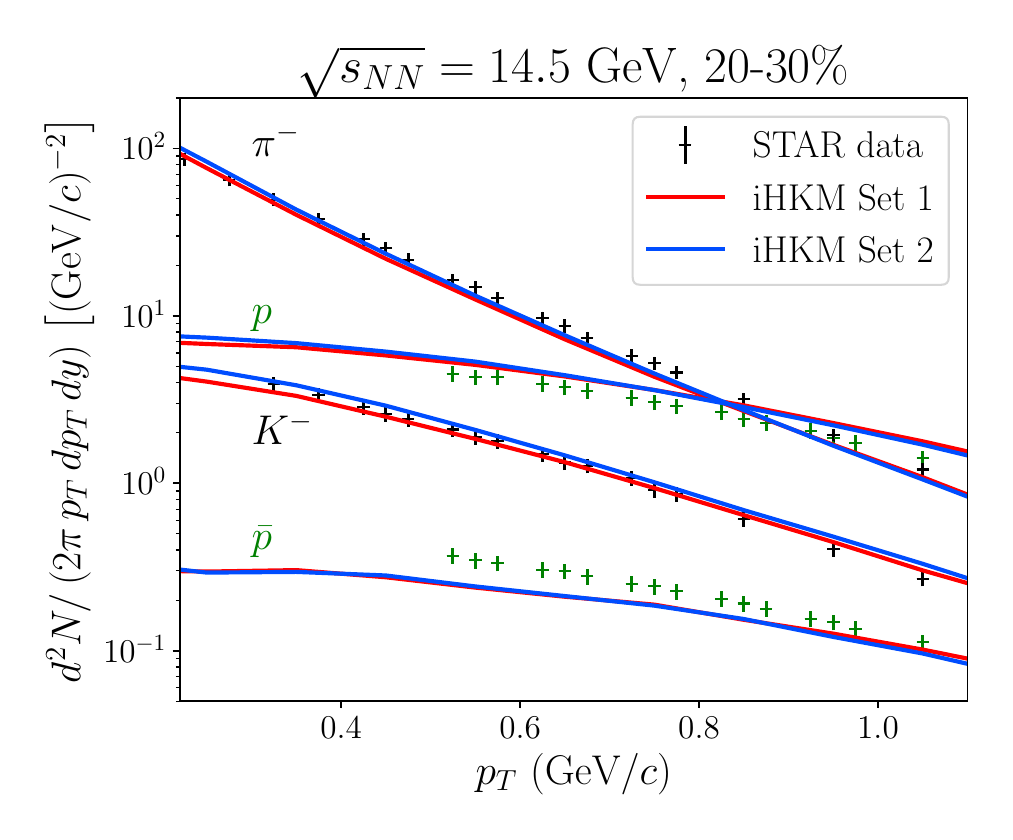}

    \caption{Transverse momentum spectra of (anti-) protons, and negatively charged kaons and pions from 0-5\% (left) and 20-30\% (right) centrality classes. iHKM parameters are described in Table~\ref{tab:optimalParams}.  STAR data is taken from  \cite{STAR:2019vcp}.}
    \label{fig:spectraMain}
\end{figure*}

This section presents our results using the tuned set of free parameters. The transverse momentum spectra for the lightest hadrons production in $0-5$\% and $20-30$\% centrality classes are shown in Fig.~\ref{fig:spectraMain}. It is evident that the model underestimates the $\bar{p}/p$ ratio, particularly noticeable in central collisions, and adjusting the free parameters does not fully resolve this discrepancy. This issue might be mitigated by considering dissipative terms in the baryon current (\ref{eq:Jtilde}), which could reduce the baryon chemical potential within the midrapidity region of the system. However, it could also indicate an overestimation of baryon stopping in the UrQMD model in combination with the hydrodynamization in the intermediate collision energy regime.
    
\begin{figure}
    \centering
    \includegraphics[width=0.5\textwidth]{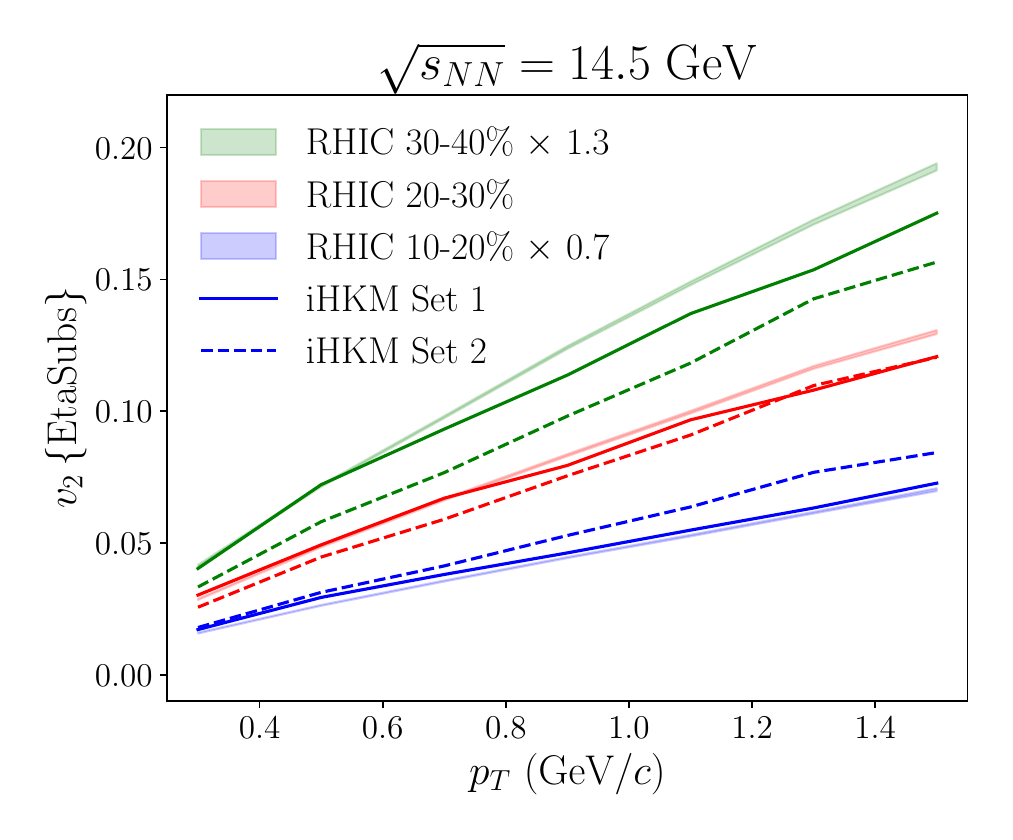}
    \caption{Elliptic flow coefficients $v_2$ dependents on transverse momenta based on $\eta$-substraction method. The experimental data is taken from \cite{STAR:2019vcp}. The dashed region accounts for statistical error. Additional multipliers were added to visually separate plots from different centrality classes. Parameters of iHKM are described in Table~\ref{tab:optimalParams}.}
    \label{fig:v2}
\end{figure}

We also compare iHKM results for $p_T$ dependence of the elliptic flow obtained via $\eta$-substraction method $v_2\{\eta-\text{Sub}\}$. Three centralities presented in STAR collaboration paper \cite{STAR:2019vcp} are considered. In Fig.~\ref{fig:v2} one can see that the chiral EoS (Set 1) results are very close to the data. Meanwhile, iHKM calculations with a phase transition (Set 2) struggle to reach the experimental value of the flow in non-central collisions even at low share viscosity to entropy ratio $\eta/s=0.08$. Further viscosity reduction improves only the high-$p_{T}$ behavior ($p_T>1$~GeV/$c$) while dramatically worsening the spectra. The model results for $v_2$ could be improved by decreasing thermalization time $\tau_{th}$, but too fast thermalization seems unrealistic. Another possibility to improve results is to treat $\tau_0$
 and $\tau_{th}$ as free parameters for different centrality classes. However, this is out of the scope of this paper.
 
\begin{figure*}
    \centering
    \includegraphics[width=0.47\textwidth]{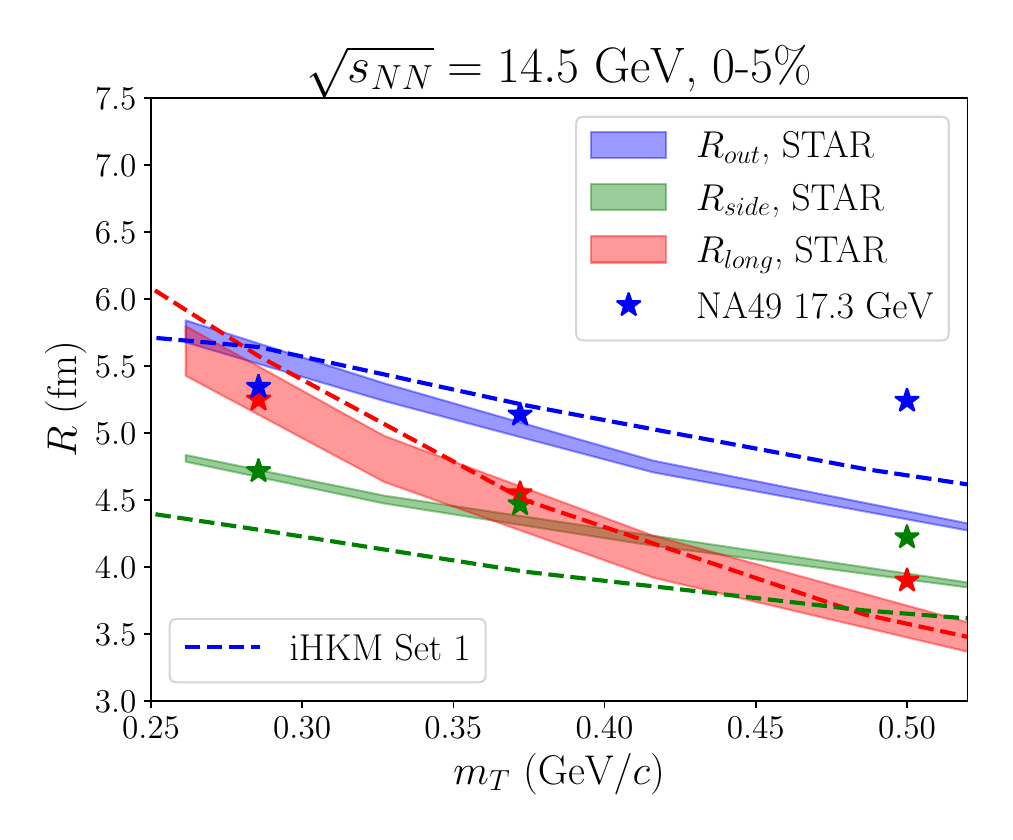}
    \includegraphics[width=0.47\textwidth]{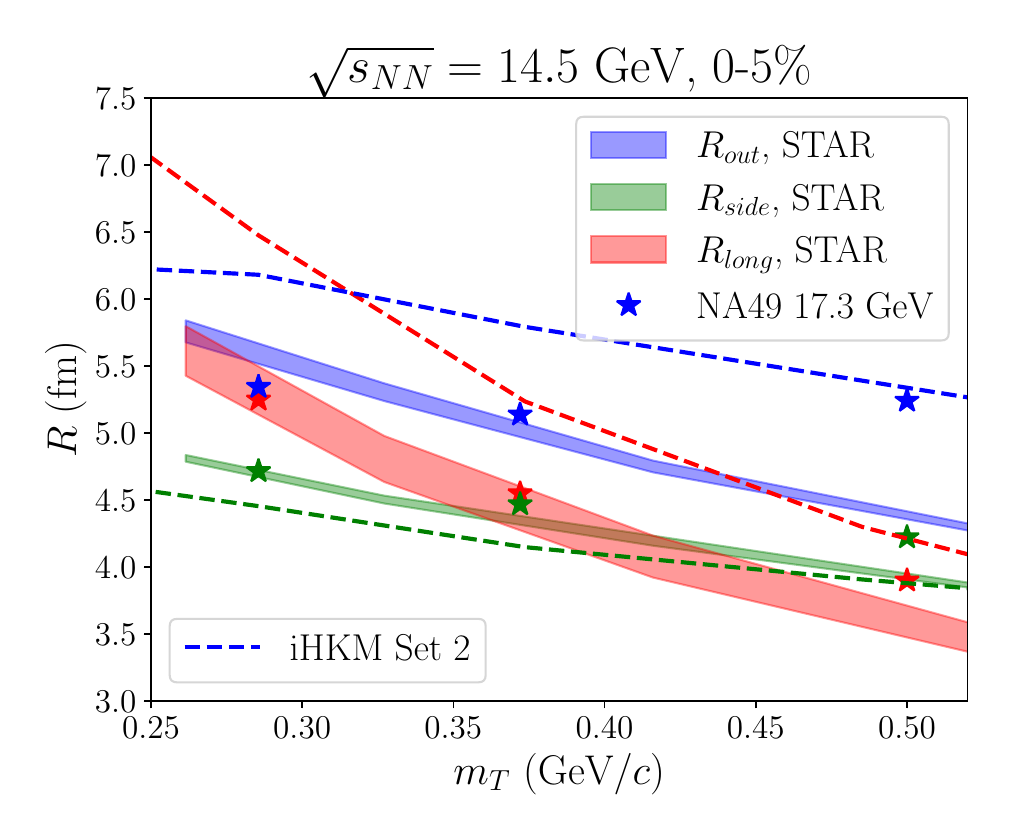}
    \caption{ iHKM results for the $m_T$ dependence of $R_{out}$, $R_{side}$ and $R_{long}$ for negatively charged pions in $5\%$ the most central collisions of gold nuclei at $\sqrt{s_{NN}}=14.5$~GeV. Left: iHKM results obtained with crossover equation of state, and right: with first-order phase transition. Shaded regions cover values between STAR measurements for $11.5$~GeV and $19.6$~GeV \cite{STAR:2014shf}. Additionally, experimental results from the NA49 collaboration for Pb+Pb collisions at $17.3$~GeV \cite{Das:2018grv} are included for reference.}
    \label{fig:hbt}
\end{figure*}

Lastly, we present our two-pion interferometry results for $5\%$ most central collisions. Particle selection was done according to STAR acceptance \cite{STAR:2014shf}: $\lvert y \rvert < 0.5$, $p_T>0.15$~GeV/$c$. Correlation functions in longitudinal co-moving system (LCMS) frame were fitted in low relative momenta region $\lvert q\rvert < 0.15$~GeV/$c$ via the thee-dimensional Gaussian function:
\begin{equation}
\begin{aligned}
C(k_T, q) &= 1 + \lambda \exp\left[ -q^2_{\text{out}}R^2_{\text{out}} \right. \\
&\quad \left. -q^2_{\text{side}}R^2_{\text{side}} -q^2_{\text{long}}R^2_{\text{long}}\right],
\end{aligned}
\end{equation}
where we use standard Bertsch-Pratt notation for \emph{out-side-long} coordinate system \cite{Bertsch:1988db, Pratt:1986cc}. Resulting $R_{out, side, ling}$ dependencies on $k_T$ presented on Fig.~\ref{fig:hbt}. Since experimental data for pion femtoscopy at $14.5$~GeV has not been published yet, we present experimental data for two neighboring energies $11.5$~GeV and $19.6$~GeV \cite{STAR:2014shf}, assuming that expected values must fall between them. 

As shown in Fig.~\ref{fig:hbt}, similar to the elliptic flow $v_2$ results, iHKM performs significantly better when the equation of state (EoS) with a crossover-type transition is used. The substantial overestimation of $R_{\text{long}}$ in model calculations with Set~2 can be attributed to the relatively long hydrodynamic stage compared to the Set~1 scenario. This difference arises from both the characteristics of a softer (lower-pressure) EoS at low temperatures and a later transition to the hadronic stage due to the lower switching energy density $\epsilon_{\text{sw}}$. 

A qualitatively similar result for the $R_{\text{long}}$ component using the same equations of state, including the hadronic EoS, was reported in Ref.~\cite{Li:2008qm}, which agrees with findings. However, quantitatively, in calculations with the crossover EoS, we achieve a better description of the ``long'' and ``out'' components, but a worse description in the ``side'' direction. A detailed beam energy scan study of pion interferometry within our model is left for future work.

\subsection{Maximal emission times estimate}
Recently, a simple method for extracting the times of maximal emission for kaons and pions has been developed. This method utilizes a combined fit of their transverse momentum spectra and the dependence of the longitudinal interferometry radii on the pair transverse momentum $k_T$ \cite{Akkelin:1995gh}. For details, we refer the reader to our previous works, where this method was applied to ultrarelativistic heavy-ion collisions \cite{Shapoval:2020nec, Sinyukov:2015kga, Shapoval:2021fqg}. Here, we present only the final analytical expressions:

\begin{equation}\label{eq:rlong_fit}
    R^2_{long}(k_T) = \tau^2\lambda^2 \left( 1+\frac{3}{2}\lambda^2\right),
\end{equation}
\begin{equation}\label{spectra_fit}
    p_0\frac{d^3N}{dp^3} \propto \exp \left( - \left( \frac{m_T}{T} + \alpha\right) \sqrt{1-\bar{v}_T^2} \right),
\end{equation}
where $T$ is the effective temperature of the freeze-out hypersurface, $m_T$ is the transverse mass of the particle pair in the LCMS frame, and
\begin{equation}
    \bar{v}_T = \frac{k_T}{m_T+\alpha T}
\end{equation}
is the transverse collective velocity at the saddle point. The parameter $\alpha$, which differs for pions and kaons, characterizes the intensity of the collective transverse flow\footnote{An infinite $\alpha$ corresponds to the absence of flow, while small $\alpha$ values indicate strong flow.}. The parameter $\lambda$ is related to the homogeneity length in the longitudinal direction in the presence of transverse flow:
\begin{equation}
    \lambda^2 = \frac{\lambda_{long}^2}{\tau^2} = \frac{T}{m_T} \sqrt{1-\bar{v}^2}.
\end{equation}

\begin{figure}
    \centering
    \includegraphics[width=0.47\textwidth]{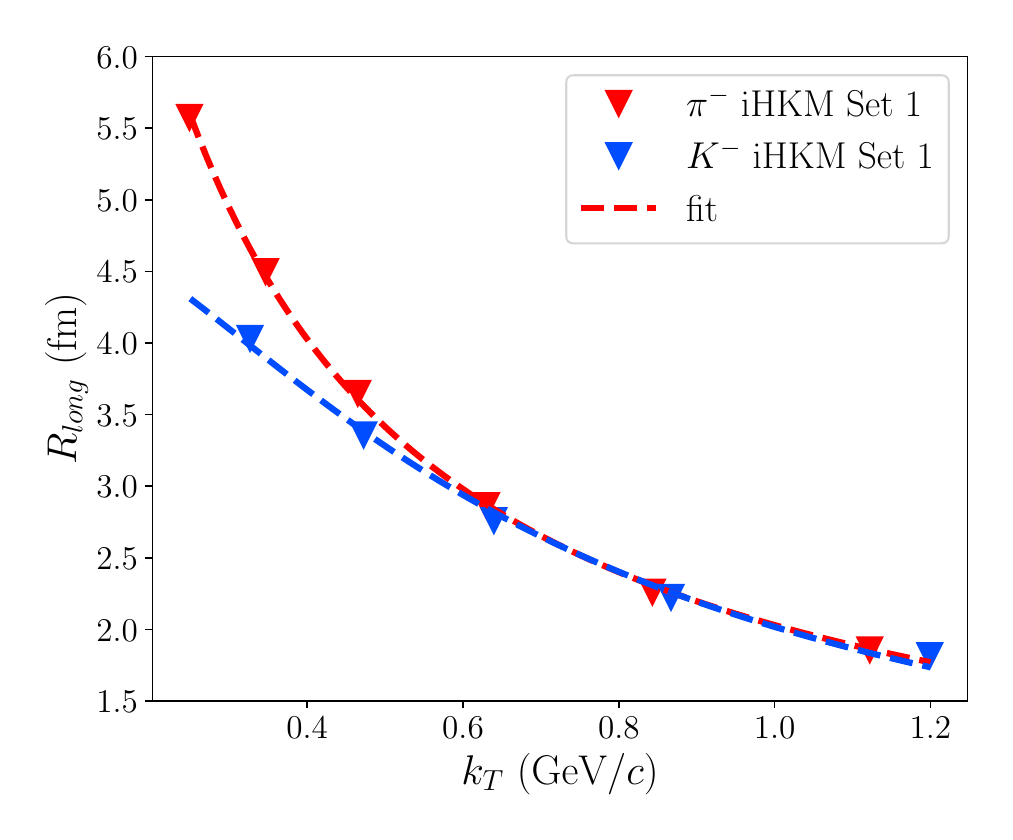}
    \caption{Transverse momentum dependence of the two-pion and two-kaon $R_{\text{long}}(k_T)$ radii obtained from the iHKM simulation with two different parameter sets (see Table~\ref{tab:optimalParams}). The dashed lines show the fits using Eq.~(\ref{eq:rlong_fit}).}

    \label{fig:hbtfit}
\end{figure}

Using Eqs.~(\ref{eq:rlong_fit}) and (\ref{spectra_fit}), we estimated the maximal emission time $\tau$ of pions and kaons in the $5\%$ most central collisions. First, we extracted the effective temperature by simultaneously fitting the pion and kaon transverse momentum spectra using Eq.~(\ref{spectra_fit}), yielding $T = 141 \pm 4.5$~GeV. Then, the remaining parameters were obtained from the $R_{long}(m_T)$ dependence using Eq.~(\ref{eq:rlong_fit}) for both scenarios considered in this study. The results are presented in Table~\ref{tab:femtoResults} and Fig.~\ref{fig:hbtfit}.

\begin{table}[b]
\centering
\caption{\label{tab:femtoResults}
Parameters obtained from the fit of transverse momentum spectra and HBT interferometry in iHKM simulations for Au+Au collisions at $\sqrt{s_{NN}}=14.5$~GeV. The temperature $T$ in both fits is  $141\pm4.5$~MeV.} 
\begin{ruledtabular}
\begin{tabular}{c|c|c|c|c}
Title & $\alpha_{\pi}$ & $\alpha_{K}$ & $\tau_{\pi}$ (fm/$c$) & $\tau_{K}$ (fm/$c$)  \\
\hline
Set 1 & $1.74\pm0.26$ & $0.24\pm 0.22$  & $6.59\pm0.13$ & $7.74\pm0.21$ \\
Set 2 & $1.82\pm0.36$ & $0.07\pm 0.06$  & $7.57\pm0.19$ & $9.18\pm0.18$  \\
\end{tabular}
\end{ruledtabular}

\end{table}

In both cases (Set~1 and Set~2), pions are emitted slightly earlier than kaons, consistent with our previous findings at ultrarelativistic energies \cite{Shapoval:2020nec}. This time difference arises primarily from the substantial decay of \( K^{*}(892) \) mesons during the afterburner stage. Additionally, the fit indicates a stronger influence of collective flow on kaon emission compared to pions, as reflected by the smaller values of \( \alpha \). This leads to \( k_T \) or \( m_T \) scaling at higher transverse momenta.

A direct analysis of the last scattering times (i.e., particle emission times) from the UrQMD afterburner confirms the consistency of these results with those reported in Ref.~\cite{Sinyukov:2023jjn}. Specifically, the maximal emission times correspond to particles with intermediate to high transverse momenta (\( 0.5 < p_T < 1.5~\mathrm{GeV}/c \)), which is the momentum range used in the fits (the region where the spectra usually exhibit exponential decay). As observed previously at LHC energies, these emission times—especially for pions—are slightly smaller than the end times of the particlization stage. For instance, in the case of the crossover equation of state (Set~1), the pion emission time is approximately \( 6.6~\mathrm{fm}/c \), while the particlization ends around \( 9~\mathrm{fm}/c \), as shown in Table~\ref{tab:femtoResults} and Fig.~\ref{fig:efraction}.

Finally, as discussed in the previous section, our calculations with different equations of state yield different system lifetimes while still reproducing the momentum spectra. This result aligns well with the generally longer lifetime of the system in the case of a softer equation of state (Set 2), which leads to larger homogeneity lengths in the \textit{long} and \textit{out} directions, as shown in Fig.~\ref{fig:hbt}.

\section{Summary}
In this work, we extend the previously developed integrated hydrokinetic model (iHKM), originally designed to describe soft physics in ultra-relativistic nucleus-nucleus collisions at top RHIC and LHC energies, to a different category of nucleus-nucleus collision experiments characterized by high central net baryon densities. Namely, those carried out at intermediate and low relativistic energies in the current BES RHIC, HADES GSI, and future CBM FAIR experiments. In both ultrarelativistic and semirelativistic cases, we are dealing with the stage of the initial formation of a state (quark-gluon or nucleon) that begins just after the overlapping of the wave packets of colliding nuclei. The possible thermalization (full or partial, depending on the collision energy) of the formed matter, and subsequent stages of the evolution of such matter are investigated. The most striking difference between the mentioned collision energy intervals is based on the time scales of collision processes. The simple estimates of the ratio of the overlapping times of wave packets at the energies per colliding nucleon pair at 5.02 TeV vs. at  7.7 GeV are about $10^{-3}$. Accordingly, the nature of the initial pre-thermal collision processes changes dramatically, particularly the time onset of thermalization of the matter evolution and its duration. These values at energies like those at BES RHIC are significantly higher than in the case of ultrarelativistic collisions.

Summarizing, a model has been developed to describe the soft physics processes at the relativistic energies  $2$ - $50$  GeV per nucleon pair. A radical modification compared to the well-known iHKM model is the simulation of the initial stage of collisions at (relatively) low energy in the quasi-classical UrQMD model instead of the CGC+GLISSANDO representation for ultra-relativistic collisions. In addition to the basic theoretical foundations of the model, we also gave examples describing within its framework the spectra of pions, kaons, protons, and antiprotons for the intermediate energy of 14.5 GeV. Publications have also been prepared to describe the spectra, elliptical fluxes, and femtoscopy radii of the mentioned particles in the energy region from 7.7 to 39 GeV/nuclear pair, describing the data and aiming to investigate the possible phase transition interval.

In this work, we developed a formalism that, in principle, can serve as a basis for the analysis of collisions at very low relativistic energies, around $2~\mathrm{GeV}$ per nucleon pair, as relevant for the GSI HADES and future FAIR CBM experiments. However, several modifications become necessary in this low beam energy regime. First, as the energy of the nuclei decreases and the nuclear overlap time increases, interactions between nucleons during the earliest stages of the collision must be taken into account. Such features are implemented in modern transport codes. Furthermore, one cannot expect significant thermalization or hydrodynamization of nuclear matter, and the conventional hydrodynamic freeze-out criterion likely loses its applicability. As discussed in this paper, this situation may require a more computationally expensive but physically motivated construction of the particlization hypersurface based on the full energy-momentum tensor.  A detailed analysis and application of our model to collisions at a few $\mathrm{GeV}$ per nucleon pair is planned for a separate publication.


\appendix

\section{Nonequilibrium distribution function in Milne coordinates}
\label{sec:appendixA}

\begin{figure*}
    \centering
    \includegraphics[width=0.47\textwidth]{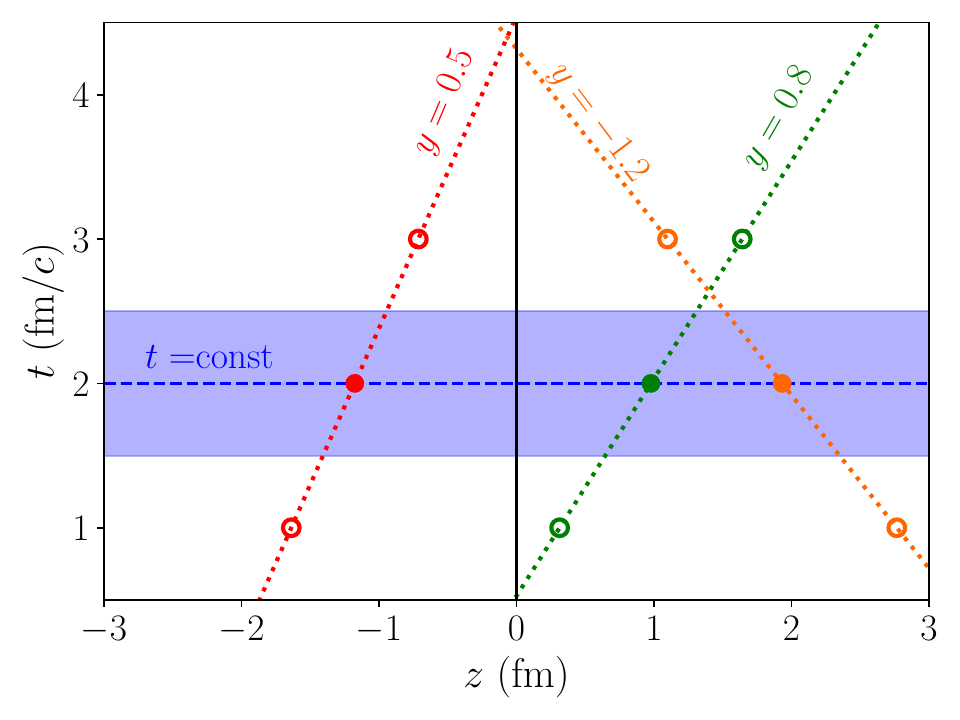}
    \includegraphics[width=0.47\textwidth]{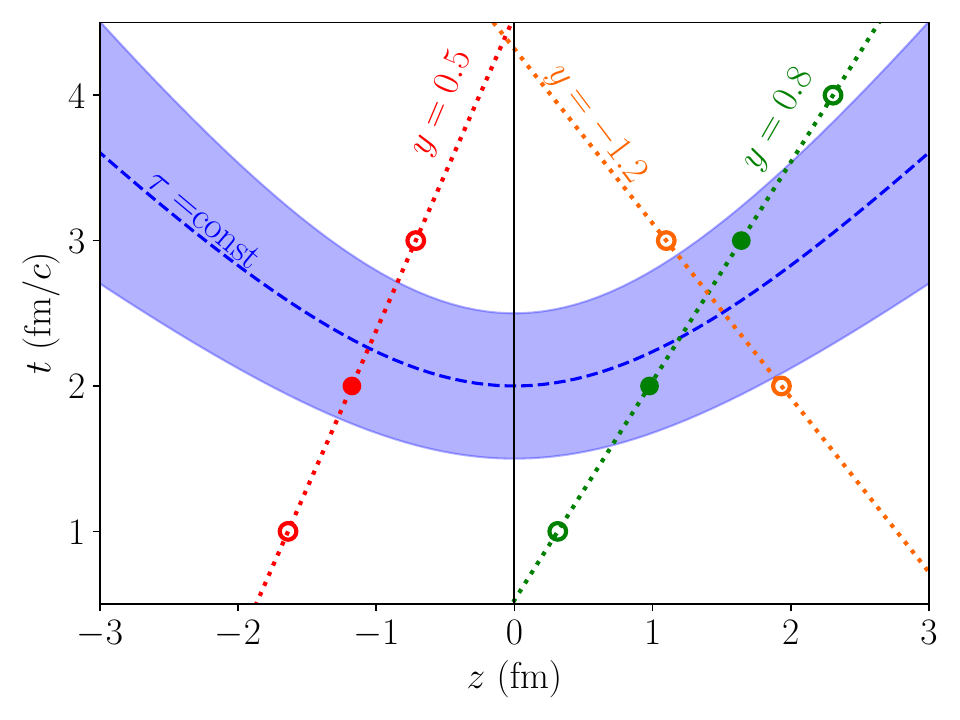}
    \caption{Tracks of three freely propagating particles crossing a layer of constant time \( t \) (left) and proper time \( \tau \) (right). The blue regions correspond to one time or proper time step \( \Delta \tau = \Delta t \). The positions of the particles are known only at fixed time steps, with \( t \propto \Delta t \). Dashed lines are added to guide the eye. Filled circles represent the parts of the tracks that lie inside the blue region, while hollow circles indicate the parts outside. In the left plot, all three particles, regardless of their initial position and rapidity \( y \), enter the blue region exactly once. In the right plot, depending on their initial position and velocity, a particle may appear zero times (orange track), once (red track), or even multiple times (green track) in the blue region.}
    \label{fig:tracks}
\end{figure*}

Let us now discuss the construction of a non-equilibrium distribution function during the relaxation stage. In this work, we utilize the UrQMD model to simulate the initial dynamics; however, this approach can be generalized to most transport models. An important point to note is that such models typically operate in Cartesian coordinates, where the system's state (the coordinates and momenta of all particles) is recorded at fixed time steps, proportional to $\Delta t$. Hydrodynamics codes, on the other hand, are generally implemented in Milne coordinates. Therefore, we require a distribution function of the system between two hypersurfaces of constant proper time, $\left[ \tau - \Delta \tau / 2, \tau + \Delta \tau / 2 \right]$. For convenience, we choose the proper time step in the hydrodynamics simulations to be the same as the time step in UrQMD, i.e., $\Delta \tau = \Delta t$.

To illustrate the consequences of transitioning to curvilinear coordinates, we consider an artificial yet realistic case, as shown in Fig.~\ref{fig:tracks}. Suppose that in UrQMD we follow the evolution of three hadrons moving freely with different rapidities. Assume that the time step is $\Delta t = 1$~fm/$c$. This value is exaggerated compared to what is used in the model, but it has been chosen for demonstrational purposes.

Now, let us consider a volume in space-time enclosed between two hypersurfaces of constant time, \( t - \Delta t/2 \leq t + \Delta t/2 \).  
As seen in the left plot of Fig.~\ref{fig:tracks}, in this region, which is painted in blue, we encounter each particle exactly once. However, if we instead consider a slice of constant proper time $\tau$ in the same range, \( \tau - \Delta \tau/2 \leq \tau + \Delta \tau/2 \), depending on the initial position and velocity of the particles, we may encounter a particle zero times (orange track), once (red track), or even two or more times (green track).

The aforementioned under- or over-counting of particles could, in principle, can lead to a violation of conservation laws. The contribution of each particle \( i \), with momentum \( p^{\mu}_i \) and mass \( m_i \), to the particle flow in the laboratory frame, when it appears in the blue region, is given by  

\begin{equation}
    N_i = \frac{n_{i\,\mu}p_i^{\mu}}{p_i^0}\,.
\end{equation}

where $n_i^{\mu}$ is the normal vector to the chosen hypersurface at the position of the particle $x^{\mu}_i$ (pointing toward the future). For the hypersurface of constant time
\begin{equation}
    n_{t}^{\mu} = \left( 1, 0, 0, 0 \right),
\end{equation}
so $N^t_i = 1$ for any particle $i$ and the conservation of flow is clear. In the case of constant proper time
\begin{equation}
    n_{\tau}^{\mu} = \left( \cosh \eta, 0, 0, \sinh \eta \right),
\end{equation}
we get 
\begin{equation}\label{eq:flowtau}
    N^{\tau}_i = \frac{\cosh\left(\eta_i - y_i\right)}{\cosh y_i}\,,
\end{equation}
which is generally not equal to one for an arbitrary particle with rapidity $y_i$ located at space-time rapidity $\eta_i$. 

This flow accounts for the under- or over-counting of particles that occurs when using curvilinear coordinates. In the toy example shown in Fig.~\ref{fig:tracks}, the values of $N^{\tau}_i$ would be close to one for the red track, less than one for the green track, and not applicable to the orange track, as that particle does not intersect the blue zone. However, if we sum $N^{\tau}_i$ over all tracks that intersect the blue region, then on average, for an arbitrary $\tau$ slice and arbitrary particle trajectories defined by $\eta_i$ and $y_i$—we should recover the total number of particles. In this example, that number is 3.

In realistic simulations, one is primarily concerned with the conservation of total energy, baryon charge, and other conserved quantities. These can be calculated using the following expressions:

\begin{equation}
    E_{\text{total}} = \sum_i p_i^0 N_i^{\tau},
\end{equation}
\begin{equation}
    B_{\text{total}} = \sum_i B_i N_i^{\tau},
\end{equation}

\noindent
where the summation is performed over all tracks appearing in the constant-\(\tau\) slice, \( p_i^0 \) is the energy of the \( i \)-th particle, and \( B_i \) is its baryon number. Our numerical simulations for RHIC Beam Energy Scan energies show that the violation of conservation laws at each proper time step \( \Delta \tau \) remains close to 1\%. This violation decreases with increasing multiplicity in more central collisions and at higher beam energies, and vanishes on average during the relaxation stage (i.e., averaged over \(\tau\)).

Now, we demonstrate that the kernel~(\ref{eq:kernelIHKM}) is properly normalized and does not lead to significant violations of conservation laws. In the continuous limit, it takes the form  

\begin{align}
    &K(x; x_i, p_i) =  \frac{ n^{\nu}_{i}u_{i\,\nu} }{(\pi R^2)^{3/2}} \times  \nonumber  \\
    &\exp\left( \frac{ (x-x_i)^2  - \left(x \cdot u_i-x_i \cdot u_i \right)^2 }{R^2}\right)\,,
\end{align}
where \( x_i \) is the space-time position of the particle, and \( u^{\mu}_i = p^{\mu}_i / m_i \) is its velocity. Then, the flow contribution (\ref{eq:flowtau}) must be adjusted as follows:

\begin{equation}
    N^{\tau}_i = \int J_i^{\mu} d\sigma_{\mu} = \int \frac{p_{i}^{\mu} d\sigma_{\mu}}{p^{0}_{i}} K(x; x_i, p_i)\,,
\end{equation}
or, explicitly,
\begin{align}\label{eq:Nsmooth}
    &N^{\tau}_i = \frac{n^{\nu}_{i} u_{i \, \nu}}{p^{0}_{i}} \int \frac{ p_{i}^{\mu} d\sigma_{\mu}}{(\pi R^2)^{3/2}} \times \nonumber\\ 
    &\exp\left( \frac{ (x - x_i)^2 - \left(x \cdot u_i - x_i \cdot u_i \right)^2 }{R^2} \right)\,.
\end{align}

The integral in this expression can be easily evaluated in the particle's rest frame, where \( p_i^{*\,\mu} = (m_i, 0, 0, 0) \), and it equals the mass of the particle, \( m_i \). Since this integral is a Lorentz scalar by construction, its value remains unchanged in any frame. Therefore, in the laboratory frame, we obtain  

\begin{equation}\label{eq:Ni}
    N^{\tau}_i = m_i \frac{n^{\nu}_{i} u_{i \, \nu}}{p^{0}_{i}} = \frac{n^{\nu}_{i} p_{i \, \nu}}{p^{0}_{i}} = \frac{\cosh \left( y_i - \eta_i \right)}{\cosh y_i}\,,
\end{equation}  
which exactly matches the expression in Eq.~(\ref{eq:flowtau}). Thus, in the continuous limit, introducing the smoothing kernel~(\ref{eq:kernelIHKM}) does not violate conservation laws. Numerically, for RHIC BES energies, we observed violations of total energy and baryon charge that usually do not exceed \( 0.5\% \), attributed to the discreteness of the lattice with spacing \( \Delta x = \Delta y = 0.3 \)~fm, \( \Delta \eta = 0.05 \), and a Gaussian radius parameter \( R \) in the range of \( 0.5 \)–\( 1.0 \)~fm.

\begin{acknowledgments}
The authors thank O. Vityuk and V. Naboka for their initial scientific contributions to this topic. They also acknowledge N. Rathod and H. Zbroszczyk for their support and valuable input in the development of the model. M.A. gratefully acknowledges support from the Simons Foundation (Grant SFI-PD-Ukraine-00014578). The research of Yu. S. was partially funded by IDUB-POB projects granted by WUT (Excellence Initiative: Research University (ID-UB)). This work is part of a project proposal submitted to the National Science Centre (OPUS 29 call, currently under review).
\end{acknowledgments}



\bibliography{apssamp}

\end{document}